\documentclass[pre,aps,twocolumn,showpacs]{revtex4-2}
\usepackage{graphicx,epstopdf}
\usepackage{dcolumn}
\usepackage{bm}
\usepackage{color}
\usepackage{xcolor}
\usepackage{float}
\usepackage{amssymb,amsmath}
\definecolor{darkgreen}{rgb}{0.0, 0.5, 0.0}
\usepackage{tabularx}
\begin{document}
\title{Fine structures of Intrinsically Disordered Proteins}

\author{Swarnadeep Seth}
\author{Brandon Stine}
\author{Aniket Bhattacharya}
\altaffiliation[]
{Author to whom the correspondence should be addressed}
{}
\email{Aniket.Bhattacharya@ucf.edu}

\affiliation{Department of Physics, University of Central Florida, Orlando, Florida 32816-2385, USA}
\date{\today}
\begin{abstract}
  \centerline{\bf ABSTRACT}
  \vskip 0.125truecm
We report simulation studies of 33 single intrinsically disordered proteins (IDPs) using coarse-grained (CG) bead-spring models where interactions among different amino acids are introduced through a hydropathy matrix and additional screened Coulomb interaction for the charged amino acid beads. Our simulation studies of two different hydropathy scales (HPS1, HPS2) [Dignon {\em et al.}, PLOS Comp. Biology, 14, 2018, Tesei {\em et al.} PNAS, 118, 2021] and the comparison with the existing experimental data indicates an optimal interaction parameter $\epsilon = 0.1$ kcal/mol and $0.2$ kcal/mol for the HPS1
and HPS2 hydropathy scales. We use these 
best-fit parameters to investigate both the universal aspects as well
as the fine structures of the individual IDPs by introducing
additional characteristics. (i) First, we investigate the polymer specific
scaling relations of the IDPs in comparison to the universal scaling
relations [Bair {\em et al.}, J. Chem. Phys. {\bf 158}, 204902 (2023)]
for the homopolymers. By studying the scaled end-to-end
distances $\langle R_N^2\rangle/(2 L\ell_p)$ and the scaled transverse
fluctuations
$\tilde{l}^2_{\perp}=\sqrt{\langle{l_{\perp}^2}\rangle}/{L}$ 
we demonstrate that IDPs are broadly 
characterized with a Flory exponent of $\nu\simeq0.56$ with the
conclusion that conformations of the IDPs interpolate between Gaussian
and self-avoiding random walk (SAW) chains.  Then we introduce (ii)
Wilson charge index ($\mathcal{W}$) that captures the essential
features of charge interactions and distribution in the sequence space
and (iii) a skewness index ($\mathcal{S}$) that captures the finer
shape variation of the gyration radii distributions as a function of
the net charge per residue and charge asymmetry parameter. Finally, our study of the (iv) variation of $\langle R_g\rangle$ as a function of salt concentration provides another important metric to bring out finer characteristics of the IDPs which may carry relevant information for the origin of life.
\end{abstract}
\maketitle
\section{Introduction}
Intrinsically disordered proteins (IDPs) are characterized by
a low proportion of hydrophobic residues
and a high content of polar and charged amino acids which make them
distinct from those which fold~\cite{Uversky2000}-\cite{Pappu2014}. IDPs lack well defined three-dimensional structures
and do not participate in forming $\alpha$-helices or $\beta$-strands,
and other secondary or tertiary structures.
Since their discovery almost three decades ago the number of IDPs has
been growing at a steady rate~\cite{DisProt}. It is now known that almost 30\% of the
proteins are either IDPs or folded proteins have intrinsically
disordered regions (IDR)~\cite{Giansanti} which play crucial roles in numerous biological processes, such as regulating
signaling pathways, helping in molecular recognition,
in initiating protein-protein interactions, and serve
as molecular switches~\cite{IDP_cell_signalling,Ferrie}. 
The conformal flexibility of IDPs helps mediate interactions with binding partners to 
form components of macromolecular complexes~\cite{Best_Nature2018,Schuler_NatCommn}. The flexibility and
often faster dynamics allow IDPs to bind to multiple different
proteins~\cite{Ferrie,Fung2018}. The IDP complexes~\cite{Best_Nature2018} have also been realized to play a central to the pathology of several degenerative diseases:
$\alpha$-synuclein (Parkinson’s disease), tau (Alzheimer’s disease), and IAPP
(Type II Diabetes)~\cite{IDP_Review4,Uversky2022}.
IDPs form membrane-less intra-cellular compartments, have been
identified to be the key drivers of liquid–liquid phase separation in
the cell~\cite{Pappu_NatPhys,Mittal_Nature,Chan_PNAS}, and therefore, have generated tremendous interest in the
underlying  phase separation in IDPs, formation of IDP complexes and
the role of charge separation in such systems using well established
concepts of polymer physics~\cite{Best_Nature2018,Pappu2010,Pappu2013,Schuler_PNAS2012}
\par
Evidently the study of IDPs in the last two decades
has been an active area in various branches of science. Despite
tremendous growth and interest in studying IDPs - the discovery of new
IDPs and their fast dynamics made it difficult to study
experimentally using small angle X-Ray scattering (SAXS)~\cite{Svergun}, single molecule
F\"{o}rster resonance energy transfer (smFRET)~\cite{Schuler_PNAS2012, Schuler_Review,Schuler_JCP2018,Best_FRET} and solution nuclear magnetic
resonance (sNMR)~\cite{Tompa2013} which have produced conflicting results. The
conformational information, such as end-to-end distances and gyration
radii, are also available for a limited number of IDPs.
Thus, an integrative structural biology approach that
combines experimental techniques~\cite{Gomes_SIC1}, using NMR spectroscopy
SAXS, smFRET, combined with computational
methods seem to be a practical and feasible approach
to unravel the conformational properties and interactions
of IDPs, shedding light on their structural ensembles.\par
Historically, computer simulation studies of CG models of polymers have played an important role
as a stand-alone discipline between theory and experiments
successfully predicting conformational and dynamic properties of
neutral and charged polymers~\cite{Sokal}. Similar studies have been
generalized for the IDPs taking into account different sizes, charges,
and hydropathy indices of the 20 different CG amino acid
beads~\cite{Ausbaugh,Mittal2018,Larsen2021,Pappu_Package,Thirumalai_2019,Thirumalai_2023}.
Data driven approaches aided by Machine Learning algorithms have
helped to build hydropathy indices for numerical studies of
IDPs~\cite{Larsen2021,Best_expt}. 
In this article we enlarge the scope of validity of a subset of these
models (HPS and M3) by comparing simulation results obtained for a
large number of IDPs using these two models with experimental
data~\cite{COINT,FhuA,IDP_Review3,OPN,Gomes_SIC1,histatin5}
and with the
results contained using other CG
models~\cite{Pappu_Package,Thirumalai_2019}.
\section{Coarse grained models of IDPs}
One of the hallmarks of IDPs is their characterization using the
Uversky plot ~\cite{Uversky2000}, where it has been shown 
that when the mean net absolute charge $\langle Q \rangle$
of a polypeptide chain at
neutral pH is plotted against the mean side chain hydropathy $\langle
H \rangle $, measured using the Kyte-Doolittle~\cite{Kyte}
hydrophobicity scale, a 
boundary line
\begin{equation}
\langle Q \rangle =
2.785\langle H \rangle - 1.151
\label{QH_plot}
\end{equation}
separates the compact (natively folded or
globular) and expanded (coil-like or pre-molten globular)
conformations~\cite{Uversky2000,Habchi}. Habchi {\em et al.}
~\cite{Habchi} improved
Eqn.~\ref{QH_plot} but the basic observation remains the same.
Evidently, a larger charge and a
smaller hydropathy ensure the extended structure of the IDPs. The simplicity has been appealing to build 
CG models of IDPs based on hydropathy, where the standard bead-spring
model of a homopolymer has been generalized to incorporate the
relative well depth between any two amino acids through a hydropathy matrix.~\cite{Ausbaugh}. 
Mittal and coworkers~\cite{Mittal2018,Mittal2022} have used this HPS model to compare the gyration
radii for several IDPs and found a reasonably good agreement. A
slightly different version has been used by Tesei~{\em et al.}~\cite{Larsen2021}. Unlike the HPS
models~\cite{Ausbaugh,Mittal2018,Larsen2021} where hydropathy is
introduced directly, other implicit solvent CG models have been used to study various properties of IDPs. Pappu and coworkers
developed a software called  ABSINTH (Assembly
of Biomolecules Studied by an Implicit, Novel, and Tunable
Hamiltonian) to study phase transitions in
IDPs~\cite{Pappu_Package}. Thirumalai and coworkers used another CG
model called SOP-IDP (self-organized polymer (SOP) coarse-grained
model for IDPs) with a finer level of granularity where,
except for Glycine and Alanine, the rest of the amino acid residues are
represented using a backbone bead and a side-chain (SC)
bead~\cite{Thirumalai_2019}. All these models are computationally more
efficient compared to the models with explicit solvent molecules and hence can be used to
study macromolecular condensates of IDPs leading to liquid-liquid phase
separation~\cite{Pappu_NatPhys,Mittal_Nature,Chan_PNAS,Alberti-LLPS2019,Dorfmann-LLPS2019, McCarty-LLPS2019,Aksimentiev_JPCL2020,Muthukumar_MM2022} in membraneless organelles.\par
Since their discovery, progressively more IDPs are
being cataloged~\cite{DisProt}. Compared to the models for the folded proteins, the
CG models of the IDPs are relatively new. Due to their flexibility and
faster dynamics, the experimental studies of the IDPs are relatively limited and often very difficult to interpret. Thus, studies of several CG models with convergence to the
experimental results is an important aspect of developing a better understanding of the fundamental physics of the IDPs  which share properties of the polyelectrolytes and polyampholytes, but exhibit very different sequence-specific behaviors.
In this article we critically examine the parameters of two hydropathy
models of the IDPs by enlarging the scope of previous
work~\cite{Mittal2018}-\cite{Larsen2021}. The gyration radii obtained
from the simulation then have been reanalyzed with reference to the
universal properties of the homopolymer model. Another important aspect
of this work is that we have introduced new physically motivated
metrics to analyze these fine structures of the IDPs. Finally, we 
provide specific examples in the result section, that IDPs clustered around different regions of the Charge-Hydropthay space may exhibit markedly different
characteristics.
\begin{figure}[ht!]
\begin{center}
\includegraphics[width=0.48\textwidth]{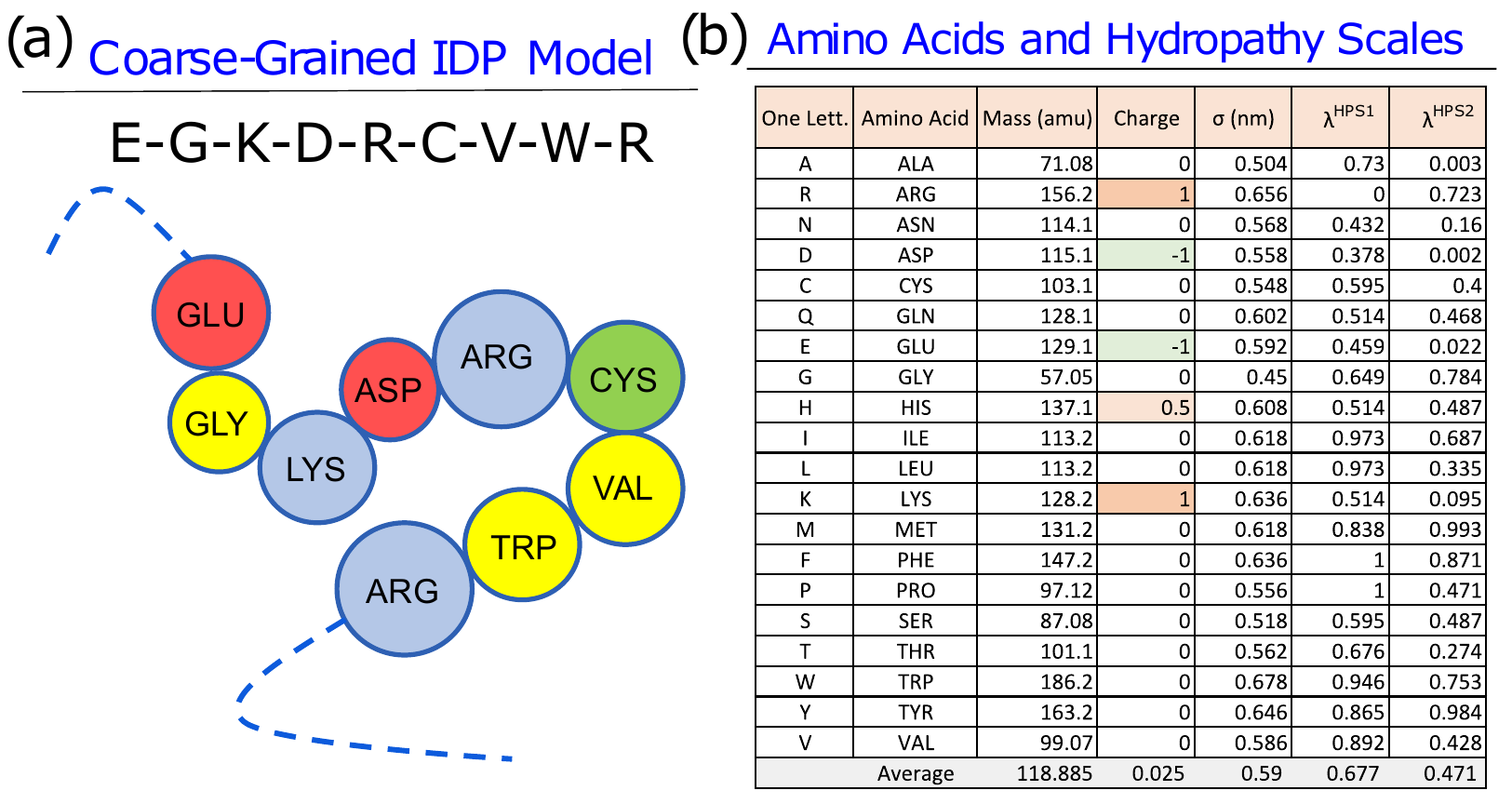}
\end{center}
\caption{\small (a) Coarse-grained model of a
  portion of IDP containing 9 amino acid residues with one letter
  code EGKDRCVWR. Each amino acid
  residue is represented by the corresponding three letter code bead of different diameter $\sigma$
  (not to the scale) listed on the table. (b) The table lists the
  one-letter and three-letter codes of the amino acids in the $1^{st}$
  and $2^{nd}$ columns. The mass, charge, and diameter of the amino
  acid residues are tabulated in the $3^{rd}$, $4^{th}$, and $5^{th}$ columns respectively. Finally, the HPS1 and HPS2 hydropathy scales are shown in $6^{th}$ and $7^{th}$ columns.}
\label{Model}
\end{figure}
\section{Simulation Model} 
The CG amino acid residues interact among themselves by a modified Van der Waals interaction potential, first introduced by Ashbaugh and Hatch~\cite{Ausbaugh} given by
\begin{align}
U_{VdW}\left(r_{ij}\right) = \begin{cases}
      U_{LJ}\left(r_{ij}\right) + (1-\lambda_{ij})\epsilon_{ij}, & \text{$r_{j} \leq 2^{\frac{1}{6}} \sigma_{ij}$}\\
      \lambda_{ij} U_{LJ}\left(r_{ij}\right), & \text{otherwise}
    \end{cases}       
\end{align}
where $U_{LJ}$ is the Lennard-Jones (LJ) potential,
\begin{align}
U_{LJ}\left(r_{ij}\right) = 4\epsilon_{ij}
  \left[\left(\frac{\sigma_{ij}}{r_{ij}}\right)^{12} -
  \left(\frac{\sigma_{ij}}{r_{ij}}\right)^6\right], 
\end{align}
$r_{ij}=\left | \vec{r}_i - \vec{r}_j \right|$ and
$\sigma_{ij}=\frac{1}{2}\left(\sigma_i+\sigma_j\right)$ are the
distance and the effective diameter 
between the amino acid beads of diameter $\sigma_i$ and $\sigma_j$ with indices $i$ and $j$ positioned at $\vec{r}_i$
and $\vec{r}_j$ respectively. 
The strength of the van der Waal interaction interaction 
$\epsilon_{ij}= \sqrt{\epsilon_i\epsilon_j} \cong \epsilon$ chosen
  to be the same, and $\lambda_{ij}=\frac{1}{2}\left(\lambda_i+\lambda_j\right)$
is the average hydropathy factor
between any two amino acids with indices $i$ and $j$. 
A harmonic bond potential
\begin{equation}
U_{b}\left(r_{ij}\right) = \frac{k_b}{2}\left(
  \frac{r_{ij}  - r^0_{ij}}{\sigma_{ij}} \right)^2.
\end{equation}
acts between two consecutive amino acid residues $i$ and
$j=i\pm1$. The spring constant $k_b$ = 8033 kJ/(mol$\cdot$nm$^2$) =
1920 kcal/(mol$\cdot$nm$^2$), and the equilibrium bond length is $
r^0_{ij}= r_0 = 0.38$~nm.
\par
A screened-Coulomb (SC) interaction acts between any two charged amino
acids
\begin{equation}
  U_{SC}\left(r_{\alpha\beta}\right) =
  \frac{q_{\alpha}q_{\beta}e^2}{4\pi \epsilon_0 \epsilon_r}
  \left( \frac{e^{-\kappa r_{\alpha \beta}}}{r_{\alpha \beta}}\right) 
\end{equation}
where the indices $\alpha$ and $\beta$ refer to the subset of the indices $i$ and
$j$ for the charged amino acids, $\epsilon_r$ is the dielectric
constant of water, and $\kappa$ is the Debye screening
length~\cite{Israel}. 
The inverse Debye length $\kappa^{-1}$ is dependent on the ionic concentration (I) and expressed as
\begin{align}
\kappa^{-1} & = \sqrt{8 \pi l_B I N_A \times 10^{-24}} 
\end{align}
where $N_{A}$ is the Avogadro's number and $l_B$ is the Bjerrum length,
\begin{align}
l_B = \frac{e^2}{4\pi \epsilon_0 \epsilon_r k_BT}.
\end{align}
At higher temperatures, the dielectric constant typically decreases, which affects the strength of the electrostatic interactions. If the dielectric constant does not account for temperature effects, the electrostatic interactions may be overestimated, leading to unrealistic protein conformations or interactions. Hence, we implement the temperature-dependent dielectric constant of water as expressed by the empirical relation~\cite{Akerlof}
\begin{align}
\epsilon_r(T) = \frac{5321}{T} + 233.76 - 0.9297T  \nonumber \\
+ 1.147\times 10^{-3} T^2  - 8.292 \times 10^{-7} T^3. \label{temp_eps}
\end{align}
\section{Hydropathy scales} 
Historically many hydropathy scales have been introduces to model the
properties of amino acids and provide a quantitative measure of the
hydrophilicity or hydrophobicity of amino acids based on their
propensity to reside in a water-soluble or water-insoluble
environment~\cite{Kyte,Engelman,Hopp, Eisenberg,Cornette}. Each of
these scales assigns a numerical value to each amino acid, reflecting
its hydrophobic or hydrophilic nature. The scores obtained from
hydropathy scales are useful in predicting protein structure and
function. Recently, specific hydropathy scales are employed to study
formation of condensates and liquid–liquid phase separation (LLPS) behavior in IDPs. Dignon~{\em et al.} proposed HPS~\cite{Mittal2018} hydropathy scale where Proline and Phenylalanine are considered to be the most hydrophobic with $\lambda^{HPS}=1$ and Arginine is the least hydrophobic with $\lambda^{HPS}=0$. All the amino acids' hydropathy are scaled to fit in the range. Later, Tesei {\em et al.} used the Bayesian parameter-learning procedure to further optimize the hydropathy values and showed M3~\cite{Larsen2021} hydropathy scale performs better to produce radius of gyration values closer to the experiments.
\subsection{HPS1 Scale}
Dignon~{\em et al.} proposed HPS~\cite{Mittal2018} hydropathy values
which are listed in the $6^{th}$ column of Fig.~\ref{Model}(b). The
original HPS model uses the constant dielectric constant of water
$\epsilon_r = 80$ for simulating the IDPs. However, in some cases,
such as, for CoINT (row 13 in Table-I), experiments are performed at a lower temperature. To account these temperature changes we used the analytic expression of temperature-dependent dielectric constant (see Eq.~\ref{temp_eps}) which makes our model slightly different from the original HPS model. However, the amino acid hydropathy scale remains identical. We denote it as the HPS1 scale in our simulation.
\subsection{HPS2 Scale}
Tesei {\em et al.} used the Bayesian parameter-learning procedure to
further optimize the hydropathy values and showed M3~\cite{Larsen2021}
hydropathy scale performs better to produce radius of gyration values
closer to the experiments. In the original M3 model the charges of the
end amino acids of an IDP are modified and the charge of the Histidine
residue is tuned as a function of the pH. However, in our simulation
we do not alter the end charges and the charge of the Histidine
residue to make a consistent comparison of the results with the other
existing models. We denote this hydropathy scale of the amino acids as
HPS2 and are shown in $7^{th}$ column of Fig.~\ref{Model}(b).
\begin{table*}
\centering 
\includegraphics[width=0.95\textwidth]{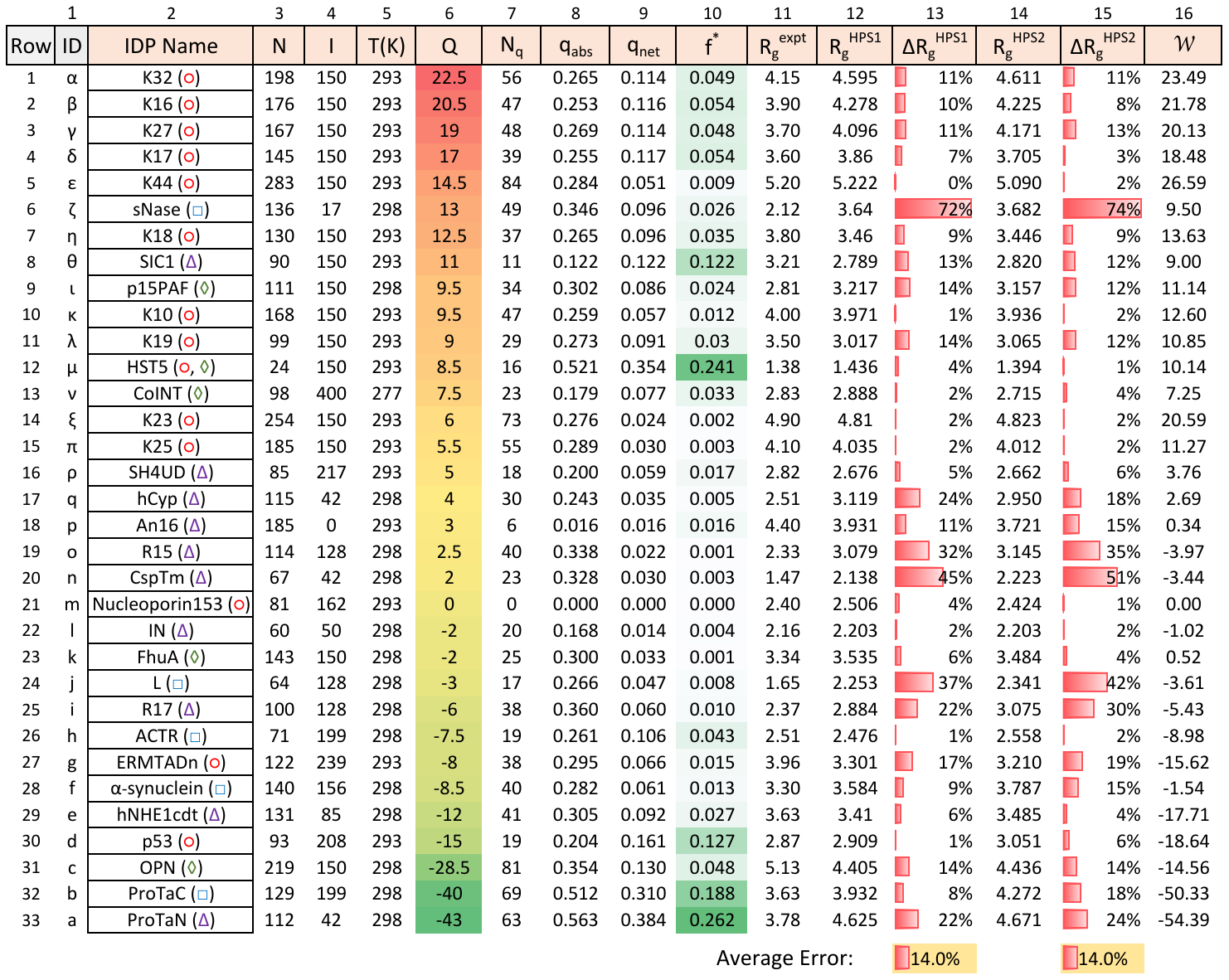}
\caption{\small Comparisons of gyration values from experimental (column 11) and  simulation studies obtained using HPS1 (column 12) and  HPS2 (column 14) models. The symbols are explained in the main text. Experimental values listed in column 11 correspond to the IDP references in the 2nd column obtained from Baul~{\em et al.}~\cite{Thirumalai_2019}({\textcolor{red}{\Large{$\circ$}}}), 
 Dignon~{\em et al.}~\cite{Mittal2018}({\textcolor{blue}{$\square$}}), 
 Tesei~{\em et al.}~\cite{Larsen2021}({\textcolor{green}{$\Diamond$}}), and Dannenhoﬀer-Lafage and Best~\cite{Best_expt}({\textcolor{violet}{$\triangle$}}). 
The 12th and 14th columns represent the simulation radius of gyration results obtained by the HPS1 and the HPS2 hydropathy scale for $\epsilon=0.1$~kcal/mol, and $0.2$~kcal/mol respectively producing least MSE (see Eqn.~\ref{MSE}, and Fig.~\ref{Rg}) with the corresponding  percentage errors in the 13th and the 15th column.}
\label{Table}
\end{table*}
\section{Results}
We studied 33 different IDPs with varying numbers of amino acids (N = 24 - 283) with
both net positive and net negative charges (see the $6^{th}$ column of
Table-~\ref{Table}). All these IDPs have been studied earlier by
different CG models~\cite{Mittal2018,Thirumalai_2019,
  Larsen2021}. However, some of the IDPs such as CspTm, Protein-L,
hCyp, R15, and R17 are either actually intrinsically disordered regions
(IDRs) of the folded proteins or unfolded state of normally folded proteins~\cite{Best_FRET,Schuler_PNAS2012}.
Within the range of our studied IDPs, An16 is a
polyelectrolyte (PE) containing only six positively charged Histidine
residues, SIC1 contains six positively charged Lysine and five
positively charged Arginine residues, Nucleoporin153 which contains
only uncharged residues, and the rest of the 30 IDPs are
polyampholytes (PAs). The table is sorted according to their net
charge $Q$ ($6^{th}$ column) from highly positive (in red) to highly negative
(in dark green) values. 
The first row of
the table lists K32 that contains 14~$\%$ Lysine (+1) and 4~$\%$
Aspartic acid (-1) which makes it highly positive. On the other hand,
the bottom row is ProTa-N that contains 20~$\%$ Glutamic acid (-1),
17~$\%$  Aspartic acid (-1) and 8~$\%$ Lysine (+1), which makes it
highly negative. We assign unique letter codes for each of the IDPs,
Greek letters $\alpha$-$\rho$ in the ascending order starting from
highly positively charged IDPs and in descending order with the alphabets
starting from the negatively charged IDPs.
The second to sixteen columns of Table-I list the names of the IDPs,
the length (N), the ionic
concentration (I) in mM, the temperature (T(K)) at Kelvin scale, total
charge $Q$, number of charged beads (N$_q$), the absolute and net charge per
unit length $q_{\rm abs}$ and  $q_{\rm net}$ (Eqn.~\ref{q_definition}),  the charge asymmetry
parameter $f^*$ (Eqn.~\ref{f*}), $R_g^{\rm expt}$, $R_g^{\rm HPS1}$,
$\Delta R_g^{\rm HPS1}$, $R_g^{\rm HPS2}$, $\Delta R_g^{\rm HPS2}$ in
units of nm, and the
Wilson parameter $\mathcal{W}$ (Fig.~\ref{Wilson_schematics}) respectively.
\par
\begin{figure*}[htb!]
\centering 
\includegraphics[width=0.95\textwidth]{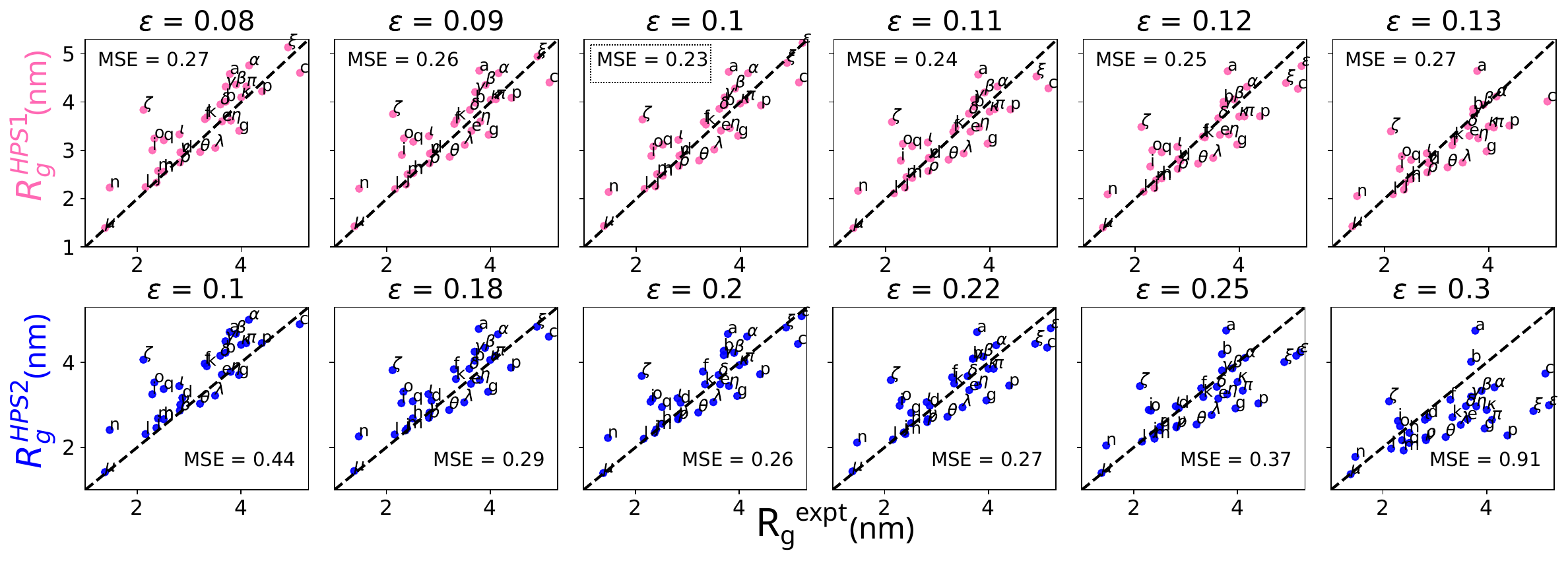}
\caption{\small The comparison of $R_g^{sim}$ with $R_g^{expt}$ for two different hydropathy models - the $1^{st}$ row shows the simulation results with HPS1 hydropathy scale (\cite{Mittal2018}) and the HPS2 hydropathy scale (\cite{Larsen2021}) results are shown in the $2^{nd}$ row for different values of LJ interaction strengths $\epsilon$ in kcal/mol unit. The black dashed line serves as a guide to show the highest positive correlation between the experiment and simulation results with a unit slope, and the deviations are calculated as mean square errors, shown as MSE, in each subplot.}
\label{Rg}
\end{figure*}
The aforementioned two hydropathy scales HPS1~\cite{Mittal2018} and
HPS2~\cite{Larsen2021} are used to study 33 IDPs using the coarse
grained simulation method described in Sec.~II. The gyration radii
$R_{g}^{HPS1}$ and $R_{g}^{HPS2}$ are obtained from
simulation using hydropathy scales HPS1 and HPS2 as a function of
$\epsilon$ (in kcal/mol) are then used to calculate the mean square error (MSE)
\begin{align}
MSE = \frac{1}{N_{tot}} \sum_{i=1}^{N_{tot}}\left[ R_g^{expt} (i) - R_{g}^{k} (i) \right]^2
\label{MSE}
\end{align}
to compare the deviation from the experimental gyration radii
$R_g^{expt}$. 
Here $N_{tot}$ is the number of different IDPs studied, and $k$
refers to either HPS1 or HPS2.
In Fig.~\ref{Rg} we show the scatter plots of the experimental
vs. gyration radii for five different
values of $\epsilon$ ($\epsilon = 0.08-0.13$ kcal/mol for the HPS1
model and $\epsilon = 0.1-0.3$ kcal/mol for the HPS2 model) and
conclude 
that $\epsilon$ = 0.1~kcal/mol and 0.2~kcal/mol
has the lowest errors (MSE = 0.23 and 0.26) for the HPS1 and HPS2
models respectively. The gyration values from these models are nearly
identical as shown in the inset (left) of Fig.~\ref{Scaling}(a).  In this paper we use HPS1 hydropathy scale with
$\epsilon$ = 0.1~kcal/mol to further analyze the properties of the
IDPs. However, we have carried out the same simulation studies using the HPS2 scale
using $\epsilon=0.2$ kcal/mol (not shown in the manuscript) and
the results are nearly identical to those from HPS1 model. Both the
scales show $\approx 14$\% error (Table-I, Columns 13 \& 15) for the set used here.
\subsection{Universal Scaling Properties of the IDPs}
\begin{figure}
\includegraphics[width=0.49\textwidth]{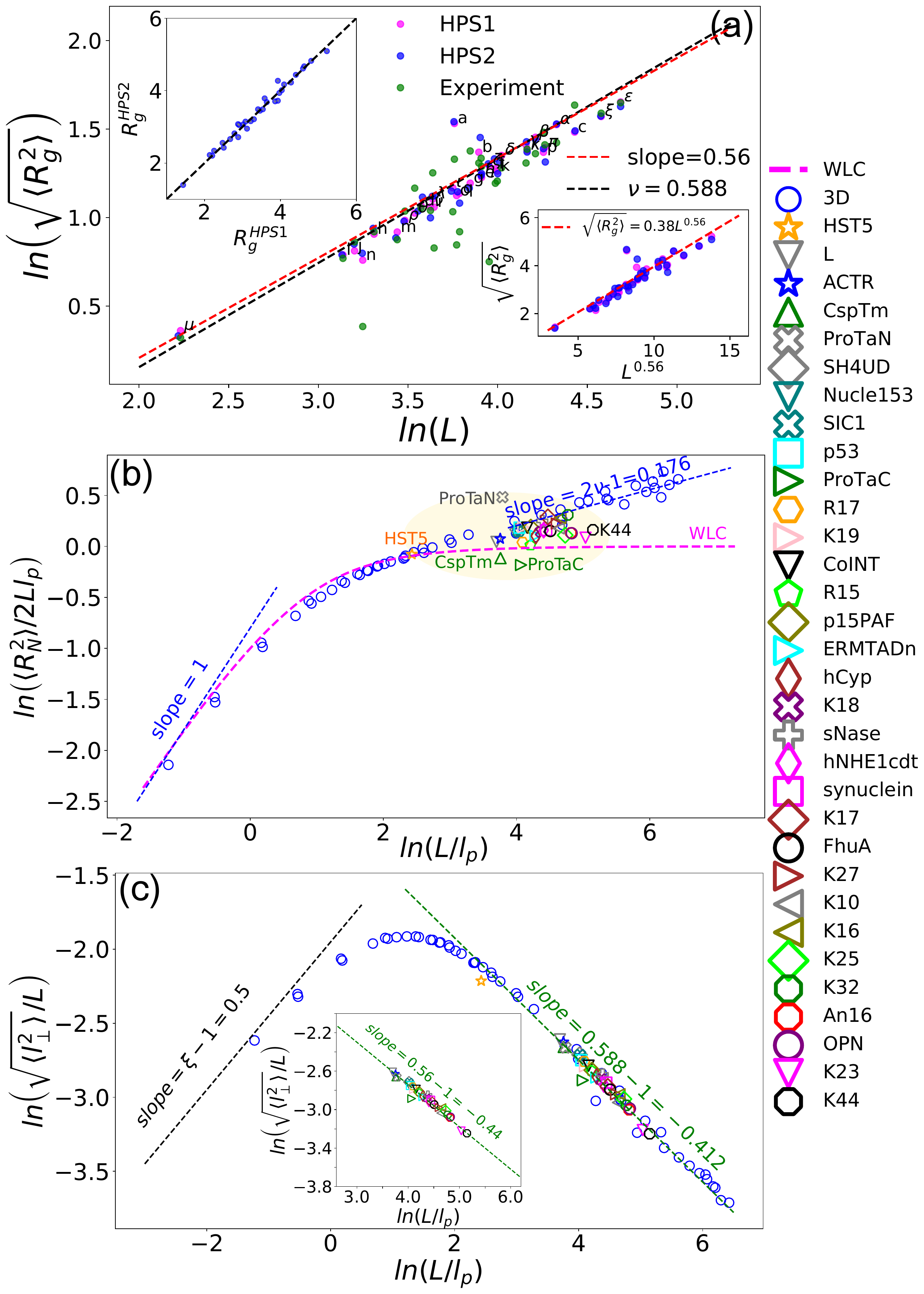}
\caption{\small (a) Log-log plot of IDPs radius of gyration as a
  function of chain length $L$ obtained from  HPS1 (magenta circles)
  and HPS2 (blue circles) models, and from the experiments green
  circles). The red dashed line shows the best fit of the simulation
  data. (top) The inset (top-left) show the comparison of the gyration
  radii from HPS1 and HPS2 models and the black dashed line indicates
  line of unit slope. The other inset (bottom-right) shows the
  simulation gyration radii  $\sqrt{\langle R_g^2 \rangle}$ as a
  function of $L^{0.56}$ and the red dashed line corresponds to
  $\sqrt{\langle R_g^2 \rangle}= 0.38L^{0.56}$  (b) Log-log plot of the scaled end-to-end distances, $\langle R_N^2 \rangle/2Ll_p$, as a function of $L/l_p$ for homopolymer chains in {\textcolor{blue}{\Large{$\circ$}}} for a variety of combinations of $L$ and $\ell_p$ for different IDPs in colored symbols for the HPS1 model. The dashed purple line in each figure shows the behavior of the WLC model [Eq.~(\ref{WLC})]. (c) Log-log plot of the scaled transverse fluctuation $\sqrt{\langle l_{\perp}^2 \rangle}/L$, as a function of $L/l_p$ with the symbols having the same meaning as (b).}
\label{Scaling}
\end{figure}

Despite the fact the IDPs are mostly described as polyampholytes (PAs) or polyelectrolytes (PEs)~\cite{Everaers_PRL}, a  fraction of experimental and theoretical studies using the HPS model by Dignon {\em et al.} describe IDPs as Gaussian chains~\cite{Mittal2022}, while  in a recent publication, Thirumalai and coworkers using a two-bead CG model calculated the RMS
$R_g\equiv \sqrt{\langle R_g^2\rangle}$ and
the end-to-end distance $R_N\equiv \sqrt{ \langle R_N^2\rangle }$, and
concluded that globally the IDPs are described not as the Gaussian
chains, but described  as fully flexible excluded volume (EV) 
self-avoiding-walk (SAW) chains that obey the Flory scaling  $R_g=
aL^{0.59}$ in three dimensions (3D), where $L$ is the contour length of the IDP~\cite{Thirumalai_2023}. We investigate this point further to find out to what extent the properties of the IDPs are universal. 
From theoretical arguments following Schaefer~{\em
  et al.}~\cite{Pincus_MM_1980} and Nakanishi~\cite{Nakanishi_1987} it
is established that a proper description of a semi-flexible EV chain characterized by a contour length $L$ and a persistence length $\ell_p$ in $d$ spatial dimensions is given by 
\begin{equation}
\sqrt {\langle R_N^2 \rangle} \simeq b_l^{\frac{d-2}{d+2}} N^{\frac{3}{d+2}}\ell_p^{\frac{1}{d+2}} = b_l^{\frac{d+1}{d+2}}\left( \frac{L}{b_l}\right)^{\nu}\ell_p^{\frac{1}{d+2}}. 
\label{Rn_EV}
\end{equation}
Here, $N$ is the number of monomers of the chain so that length of the IDP is $L = (N-1)b_l \simeq Nb_l$ (for $N \gg 1 $), $b_l$ is the bond length between two
neighboring monomers, and the mean-field Flory exponent $\nu =
3/(d+2)$ in 2D = 0.75 and in 3D = 0.60 ($\approx 0.588$ actual) respectively. 
This EV chain accurately describes the limit $L/\ell_p >> 1$ and supersedes the Worm-like-chain model~\cite{Rubinstein} 
\begin{equation}
\frac{\langle R_N^2\rangle}{L^2} = \frac{2\ell_p}{L}\left(1-\frac{\ell_p}{L}[1-\exp(-L/\ell_p)]\right).
\label{WLC}
\end{equation}
which does not take into account the EV effect and hence saturates to $\langle R_N^2 \rangle = 2 L\ell_p)$ even when $L/\ell_p >> 1$. In a previous publication 
we have shown that scaled end-to-end distance $\langle R_N^2 \rangle/(2 L\ell_p)$ and
 the scaled transverse fluctuation $\sqrt{\langle{l_{\perp}^2}\rangle}/{L}$ as a
 function of $L/\ell_p$ collapse onto the same master curve~\cite{Universal1,Universal2} for all ratios of $L/\ell_p$ spanning rod to Gaussian and the EV limit. We would like to  discuss our simulation findings for the IDPs in the context of these universal scaling plots (Fig.~\ref{Scaling}).
\par
Fig.~\ref{Scaling}(a) summarizes the simulation results from HPS1
(magenta solid circles) and HPS2 (blue solid circles) models.
For comparison we have also included the experimental
points (green circles). The red dashed line corresponds to 
$\langle R_g\rangle = 0.38L^{0.56}$ (in nm shown explicitly as a power-law
at the inset (bottom-right)). Please note that the prefactor 0.38 nm is
exactly the same as the average bond-length obtained from the simulation.
The dashed line
(black) $\langle R_g\rangle \approx L^{0.588}$  corresponds to the
excluded volume chain in 3D. 
\par
To get a clearer perspective we have calculated the length of the IDPs
and compared the scaled end-to-end distances $\langle
R_n^2\rangle/2L\ell_p$ in units of dimensionless length
$L/\ell_p$~\cite{Universal1,Universal2} (blue circles) in
Fig.~\ref{Scaling}(b). These data points will serve as a guide and
help readers to visualize the deviation of the scaling properties of
the IDPs in reference to the universal master plot for the
self-avoiding chain. 
The data points for the IDPs show that while H5T5, CspTm and ProTaC
lie in the Gaussian regime, the rest of the IDPs are in between Gaussian (WLC) and 
self-avoiding-walks in 3D but closer to being represented as
self-avoiding-walks in 3D, consistent with the conclusion from
Fig.~\ref{Scaling}(a). It is worth mentioning that a slightly lower
exponent ($0.56 < 0.588$) may be attributed partly due to the finite size effect for the chain
lengths $24 \le N \le 283$ used in this study. 
Our conclusion is further strengthened by Fig.~\ref{Scaling}(c) inset
plot, where we find that the scaled transverse fluctuations collapse
with a slope of $0.56-1=-0.44$.
These results possibly bridge and explain both the previous findings
by Dignon {\em et al.}~\cite{Mittal2022} where they identified the
Gaussian behavior of the IDPs
and by Baul {\em et al.}~\cite{Thirumalai_2019} where they described
IDPs as self-avoiding chains. 
We will see in section V-C that IDPs can be further classified in
terms of their skewness indices ($\mathcal{S}$-index captures the overall
shape).  $\mathcal{S}$-index will also affect an IDP's  global positioning with
respect to the universal master plot. For example, CspTm and ProTa-C having larger
$\mathcal{S}$-index lie on the WLC line. This finer
distinctions make IDPs more interesting. One can use the universal
curve to understand why they are more Gaussian or behave more like a
self-avoiding chain. In the next two sections we will introduce such
characteristics to further classify IDPs.\par
\subsection{Wilson index $\cal{W}$  of the IDPs}
\begin{figure}[ht!]
\begin{center}
\includegraphics[width=0.48\textwidth]{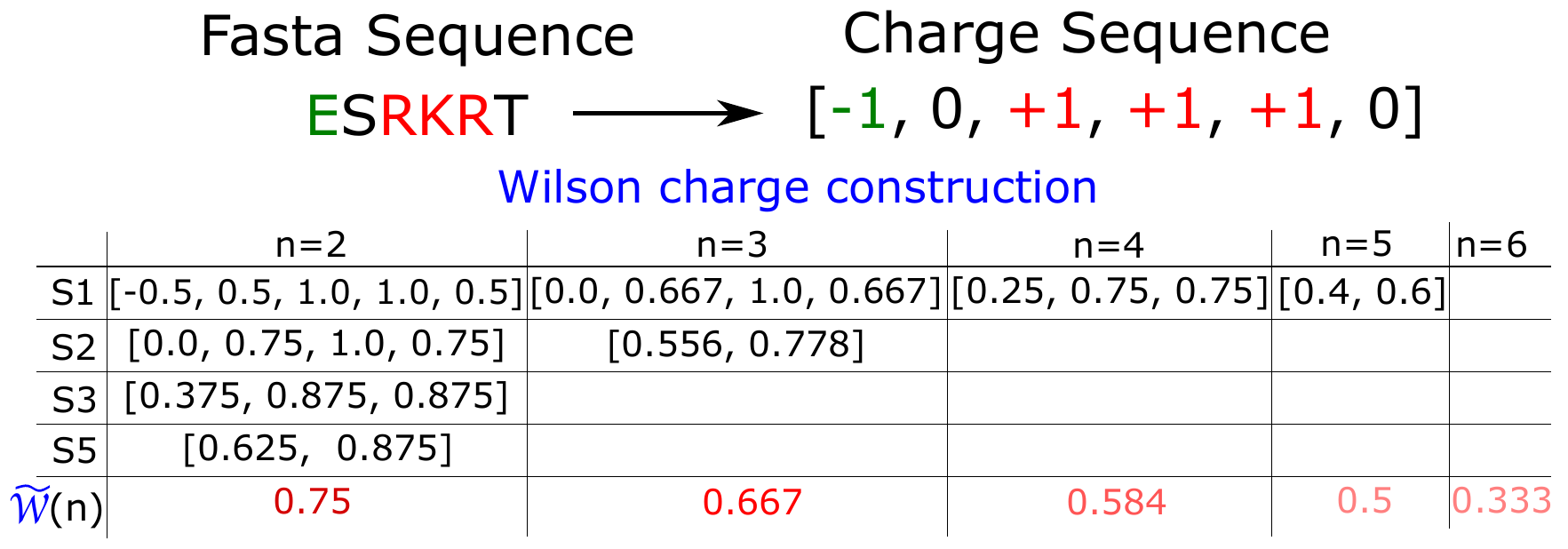}
\end{center}
\caption{\small The schematic diagram shows the derivation of Wilson charge $\tilde{\mathcal{W}}(n)$ of a hypothetical fasta sequence ESRKRT of length 6. The table shows the different $\tilde{\mathcal{W}}(n)=0.75\rm{-}0.333$ for different window lengths $n=2\rm{-}6$ averages.}
\label{Wilson_schematics}
\end{figure}
Now that we have demonstrated an approximate universal description of
the IDPs in terms of the HPS1 model, we want to demonstrate the
sequence specific features that make each IDP distinct and may
exhibit very different behaviors from their homopolymer
counterparts. Evidently, the charges present along the sequence play a
crucial role in shaping the structure and dynamics of the IDPs. Out of
20 amino acids, only five of them carry a charge. Specifically, in the
HPS1 model, both ``R" and ``K" have a $+1$ charge, ``H'' possesses a $+0.5$ charge, and ``D'' and ``E'' have charge of $-1$. However, the charges 
 are randomly distributed along the chain backbone and therefore, IDPs
 in general can be classified either as PA or
 PE~\cite{Pappu2013,Pappu2014}. Das and Pappu showed that weak
 polyampholytes form globules, whereas for a
strong polyampholyte the net charge
per residue and their linear distribution along the chain backbone
controls their conformational preferences~\cite{Pappu2013}.
FRET experiments have explicitly demonstrated expansion of charged
IDPs at low ionic concentration~\cite{Schuler2010}. Extensive research has been conducted to study this phenomenon in the existing literature. Nonetheless, our study delves deeper to investigate the positional implications of the amino acid sequence in terms of Wilson index ($\mathcal{W}$) as described below.
\par
Unlike a homopolymer, an IDP can have varied degree of local stiffness
and flexibility resulting in the amino acids in different segments interacting with neighboring sequences even if they are far apart in the sequence space.
To capture these potential dynamical interactions, we employ the concept of Wilson Renormalization extensively used to study
the spin systems~\cite{SK-Ma}. This renormalization approach allows us to analyze the sequence of charges and their unfolding interactions, considering interactions up to the next nearest neighbor. Fig.~\ref{Wilson_schematics} illustrates a hypothetical example of a short IDP sequence ``ESRKRT'' of length 6, showcasing the presence of a negative charge at the beginning followed by three positive charges in the middle, and the remaining amino acids being neutral. To initiate the averaging procedure, we select a window of length $n$. The simplest case $\tilde{\mathcal{W}}(2)$, considers sliding averages of window length n=2 and denotes the next neighbor interactions. The window length for averaging can range from 2 to $N$, where $N$ represents the number of amino acids in the IDP sequence.
For the general case of $\tilde{\mathcal{W}}(n)$, where n consecutive charges are averaged, the process begins by sliding an averaging window from one end of the sequence toward the other. After the first step of averaging, denoted as S1, we obtain a new sequence of length $N-n+1$, and use the new sequence to carry on the averaging procedure as, 
\begin{figure}[ht!]
\begin{center}
\includegraphics[width=0.45\textwidth]{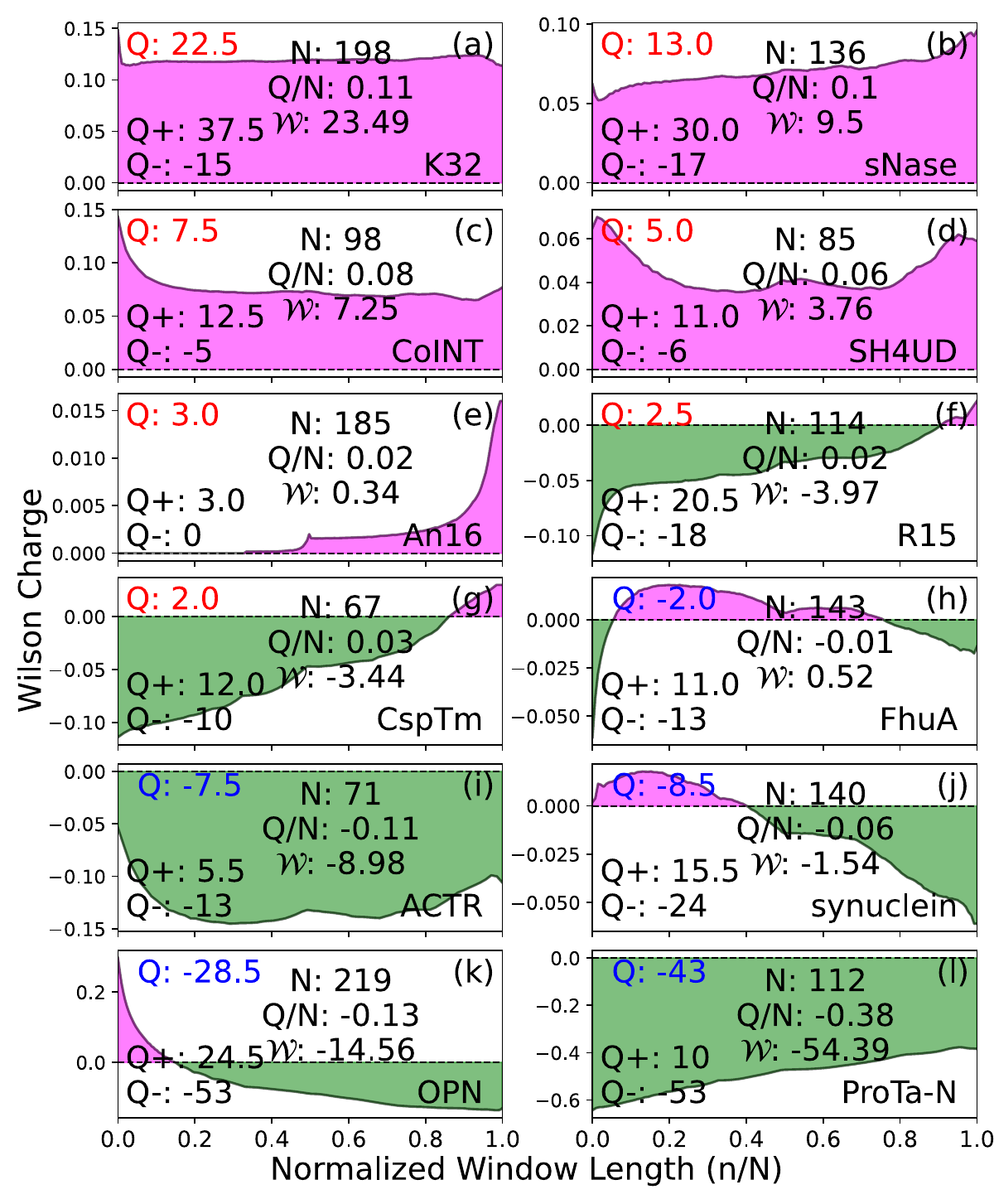}
\end{center}
\caption{\small The Wilson construction of charge distribution plotted against the corresponding window length for (a) K32, (b) sNase (c) CoINT (d) SH4UD, (e) An16, (f) R15, (g) CspTm, (h) FhuA, (i) ACTR, (j) $\alpha$-synuclein, (k) OPN, and (l) ProTa-N, in the order of positive to negative net charge content of the IDPs. $Q$ denotes the net charge of the IDP along with positive and negative charges as $Q+$ and $Q-$ respectively. $\mathcal{W}$ denotes the area under the Wilson constructed curve.  The color shade magenta/green shows the positive/negative intensity of the Wilson charge as a function of window length.}
\label{wilson_chg}
\end{figure}
\begin{subequations}
\begin{equation} 
S0:  \left[a_1, a_2, \dots , a_N \right] 
\end{equation} 
\begin{equation} 
S1:  \left[\frac{1}{n}\sum_{i=1}^{n} a_i,\; \frac{1}{n}\sum_{i=2}^{n+1} a_i, \dots, \; \frac{1}{n}\sum_{i=N-n+1}^{N} a_i\right]  
\end{equation} 
\begin{equation} 
SN:  \tilde{\mathcal{W}}(n) = \frac{1}{n}\sum_{i=1}^{n} a_i.
\end{equation} 
\end{subequations} 
We continue this procedure iteratively until we reach the final average value, represented as $\tilde{\mathcal{W}}(n)$. If the length of the charge sequence becomes less than the window length $n$ during the averaging process, we terminate the procedure and calculate a global average to obtain $\tilde{\mathcal{W}}(n)$.
\par
These averaging procedure with varied window size $n \in [2,N]$ can effectively capture the combination of charge interactions at different length scale. One can show that $\tilde{\mathcal{W}}(2)$ consider binomial interactions among the charges and expressed as 
\begin{align}
\tilde{\mathcal{W}}(n=2) = \frac{1}{2^{N-1}} \sum_{m=0}^{N-1} \binom{N-1}m a_m.
\end{align} 
The higher-order window averaging considers interactions of varying magnitudes, which can have an impact on determining dynamic conformations of IDPs. In Fig.~\ref{wilson_chg} we explore $\tilde{\mathcal{W}}(n)$ for 12 IDPs with different total charges from highly positive (a) K32 (Q=22.5), (g) CspTm (Q=2.0) to highly negatively charged IDP (l) ProTa-N (Q=-43.0) as a function of normalized window length n/N. In the case of highly positively and negatively charged IDPs, the Wilson curves consistently remain above or below the zero line respectively. However, for IDPs with lower net charges, we sometimes observe the Wilson curve crossing from negative to positive in the case of (f) R15, (g) CspTm, and from positive to negative in the case of (j) $\alpha$-synuclein, and (k) OPN. The area under the Wilson curve is denoted by $\mathcal{W}=\sum_{n=2}^{N} \tilde{\mathcal{W}}(n)$ and listed in the 16$^{th}$ column of Table-\ref{Table}. It is conceivable that when plotted in normalized unit length scale, IDPs with similar Wilson charge $\tilde{\mathcal{W}}(n)$ will behave the same way and thus, can be used as their fingerprints.\par
\subsection{Charge patches and the local persistence length}
\begin{figure*}
\centering  
\includegraphics[width=0.95\textwidth]{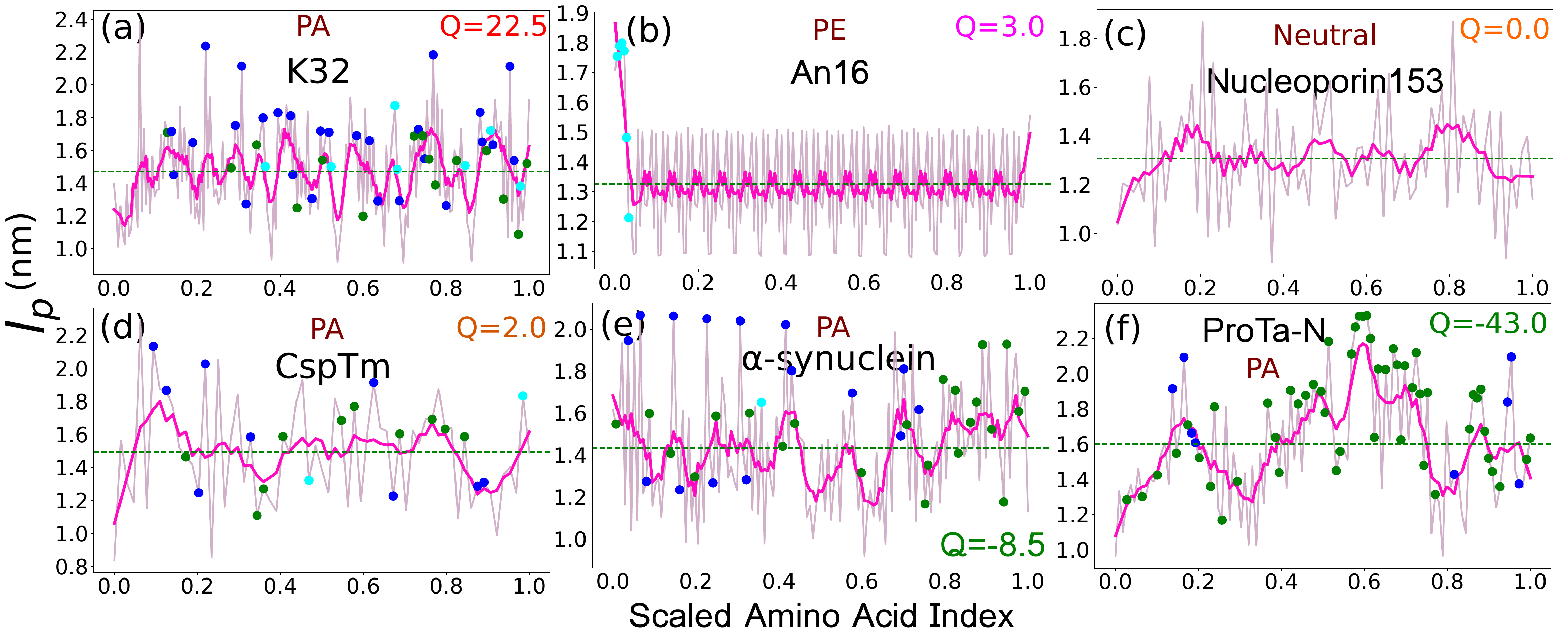}
\caption{\small The persistence length $l_p$ is shown as a function of the normalized amino acid index for (a) K32, (b) An16, (c) Nucleoporin153, (d) CspTm, (e) $\alpha$-synuclein, and (f) ProTa-N. In each case, the light pink line represents the persistence length from Eqn.~\ref{lp_sim} while the magenta line corresponds to the sliding average  
$\langle l_p \rangle = \left(\sum_{i-2}^{i+2}l_p(i)\right)/5$. The positions of charged residues along the chain backbone are indicated by filled blue {\textcolor{blue}{$\bullet$}}, cyan {\textcolor{cyan}{$\bullet$}}, and darkgreen {\textcolor{darkgreen}{$\bullet$}} circles for $+1$, $+0.5$, and for $-1.0$ charged residues respectively. The green dashed lines denote the average persistence length in each case.}
\label{lp_IDP}
\end{figure*}
 The presence of charge patches introduce varying degree of local stiffness along the chain backbone. During the BD simulation 
use a discrete chain and the persistence length is calculated from~\cite{Rubinstein}
\begin{equation}
  \ell_p/\sigma = -\frac{1}{\ln\left(\cos\theta_i\right)},
\label{lp_sim}
  \end{equation}
  where $\theta_i$ is the angle between two bond vectors connecting the $i^{th}$ bead to the $(i\pm1)^{th}$ beads~\cite{Universal2}. We have checked that for a homopolymer chain this matches well with the continuum description of persistence length~\cite{Landau}
\begin{equation}
\ell_p/\sigma = \kappa/k_BT \quad {\rm (3D)}.
    \label{lp_3d}
  \end{equation}
IDPs with very similar net charges can have markedly different distribution of charges. An IDP containing correlated charge patches will have increased chain stiffness along that region that will affect its conformations and dynamics.
To demonstrate this, we calculate the local persistence length ($l_p$) in nm along the chain using Eqn.~\ref{lp_sim} for a few IDPs shown in
Fig.~\ref{lp_IDP}. For example, CspTm has sparsely distributed charged
residues with less net charge compared to ProTa-N containing mostly negatively charged residues in patches, and we observe increase in $l_p$ on those regions. The electrostatic repulsion among the same charge residues makes the chain locally stiffer and possibly has a deeper effect in their participation in  biophysical processes.
\subsection{Skewness index ($\mathcal{S}$-index) of the radius of gyration}
\begin{figure}
\centering
\includegraphics[width=0.45\textwidth]{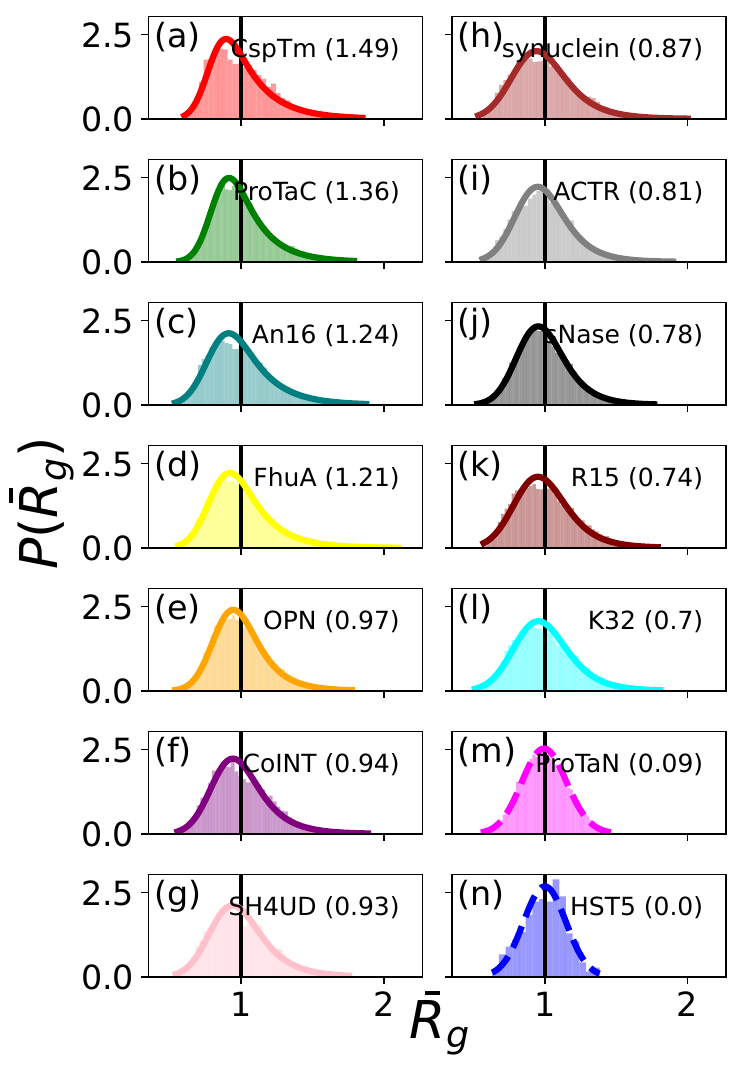}
\caption{\small $P\left(\bar{R}_g\right)$ for several IDPs as a
  function of the  $\mathcal{S}$-index. In each figure, the
  solid colored line denotes the exponentially modified Gaussian fit
  for the $P\left(\bar{R}_g\right)$ histograms (please see text) with
  the solid lines for $\mathcal{S} \ge 0.1$ while the dotted line for
  the near Gaussian distribution ($\mathcal{S} \le 0.1$). The
  corresponding $\mathcal{S}$-index are written under parentheses in the legends.}
\label{skewness}
\end{figure}

The variation of the chain persistence length due to different charge
species along the chain backbone is manifested in the shapes of the
corresponding gyration radii that we measure in terms of the 
$\mathcal{S}$-index.  The $\mathcal{S}$-index is
obtained by fitting the distribution $P\left(\bar{R}_g\right)$ of the scaled gyration
radii  $\bar{R}_g= \sqrt{R_g^2}/ \sqrt{\langle R_g^2\rangle} $ with an 
exponentially-modified Gaussian distribution
(exponnorm)~\cite{exponnorm} function given by 
\begin{equation}
f(x,K) = \frac{1}{2K}\exp\left(\frac{1}{2K^2}-\frac{x}{K}
\right){\rm{erfc}}\left(-\frac{x-1/K}{\sqrt{2}} \right)
\label{expnorm}
\end{equation}
where $x$ is a real number, $K > 0$ and erfc is the
complementary error function. 
The $\mathcal{S}$-index is then obtained from the following equation
\begin{equation}
  \mathcal{S} = \frac{2K^3}{\left(1+K^2\right)^{3/2}},
 \label{S}
\end{equation}
where $K$ corresponds to the shape parameter of the exponnorm distribution of Eqn.~\ref{expnorm}. As
shown below, the usefulness of the $\mathcal{S}$-index is
manifested in capturing the overall shape of the distributions of the the gyration
radii which then can be further analyzed as a function of the amino
acid compositions and their net charge content. We observe in Fig.~\ref{skewness} that the shapes of
the distribution of the gyration radii vary from being near Gaussian
($\mathcal{S} \rightarrow 0$) to an exponentially modified Gaussian
distribution (that exhibits a tail for larger value of $\mathcal{S}$).
\begin{figure}
\centering
\includegraphics[width=0.5\textwidth]{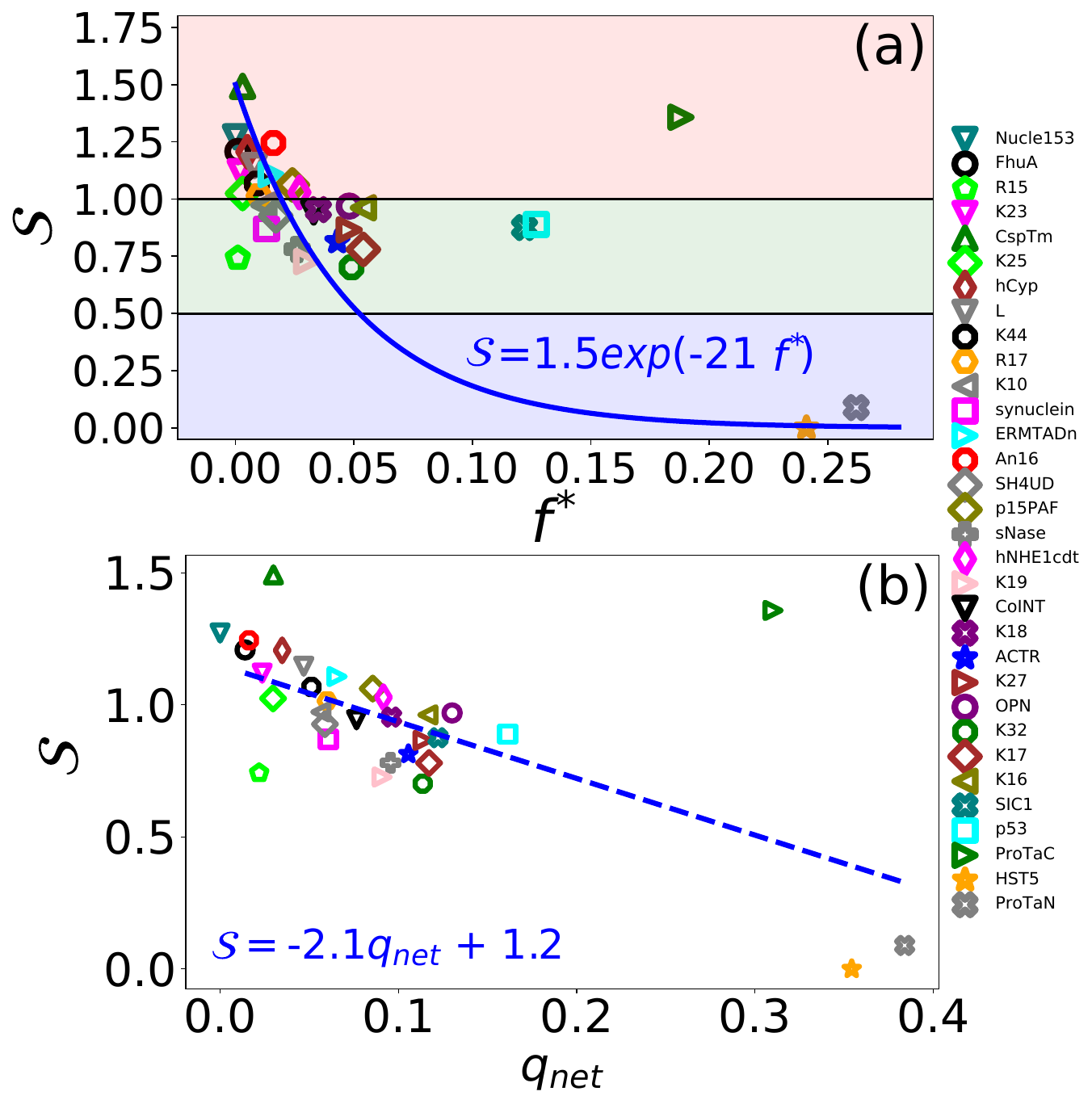}
\caption{\small (a) $\mathcal{S}$-index as a function of
  $f^{*}$. The solid blue line is an exponential fit. Light pink,
  green, and blue shaded regions denote the high, moderate, and low
  skewed radius of gyration of the IDPs. (b) $f^*$ as a function of
  net charge. The straight line is a linear fit through the points.}
\label{skewness-fit}
\end{figure}
Fig.~\ref{skewness} confirms that most of the IDPs have long
exponential tails ($\cal{S}\ge$ 1.0) such as CspTm, ProTa-C, An16,
FhuA, OPN, CoINT, SH4UD, $\alpha$-synuclein, ACTR, sNase, R15, and
K32. The highly charged IDP ProTa-N and HST5 are observed to have
almost perfect Gaussian distribution with $\mathcal{S} \rightarrow 0$. Moreover,
for the highly skewed distributions the peaks shift towards the left
that signifies the median is smaller than the mean and there is a
propensity of these IDPs to expand occasionally. This skewness
parameter can be utilized to further classify IDPs broadly into three
categories (three colored regions in Fig~\ref{skewness-fit}(a)) that characterize the propensity of expansions.
\par
We now study how the $\mathcal{S}$-index depends on the charge
content of the IDPs in more detail. In a previous study, Pappu {\em et al.}~\cite{Pappu2010,Pappu2013} demonstrated that radius of gyration depends on the charge asymmetry parameter
\begin{equation}
f^{*} = \frac{(f^+-f^-)^2}{f^++f^-},
\label{f*}
\end{equation}
where $f^+/f^-$ is the net positive/negative charge per residue of an
IDP. In Fig.~\ref{skewness-fit}(a) we plotted $\mathcal{S}$-index as a function
of the $f^*$ and based on the data it appears that there is an
exponential decay of  $\mathcal{S}$-index, although a power law fit could
not be excluded. However, when plotted as a function of the net charge
$q_{net}$ (column 9 of Table-I) 
a simpler relation is obtained (Fig.~\ref{skewness-fit}(b)). The
simulation data indicates that $\mathcal{S}$-index decreases linearly with
$q_{net}$. The linear fit which can be simplified to
\begin{equation}
  \mathcal{S} \approx -2q_{net} + 1
  \label{S-qnet}
\end{equation}
can serve as a useful relation. Since
$\mathcal{S}$ can not be negative  one observes a general relation
\begin{equation}
\lim_{q_{net} \rightarrow 0.5}\mathcal{S} \rightarrow 0\;{\rm (Gaussian)}.
\end{equation}
From Fig.~\ref{skewness} (last two rows of the right column) one observes that the shapes of the  IDPs
Prota-N and HST5 with $q_{net} = 0.384$ and 0.354 
are almost Gaussian with $\mathcal{S} = 0.09$ and $0$ respectively. This
general 
conclusion will get tested when data for more IDPs will be available and 
analyzed in future.
\subsection{Chain conformations and Ionic Concentration}
Solvent conditions, such as pH, temperature,
ionic strength, and the presence of specific molecules, can
significantly influence the conformations and hence the behaviors of
the IDPs, particularly in a cellular environment. The robustness of
the IDPs under external conditions can also be associated with the
\begin{figure*}
\centering
\includegraphics[width=0.95\textwidth]{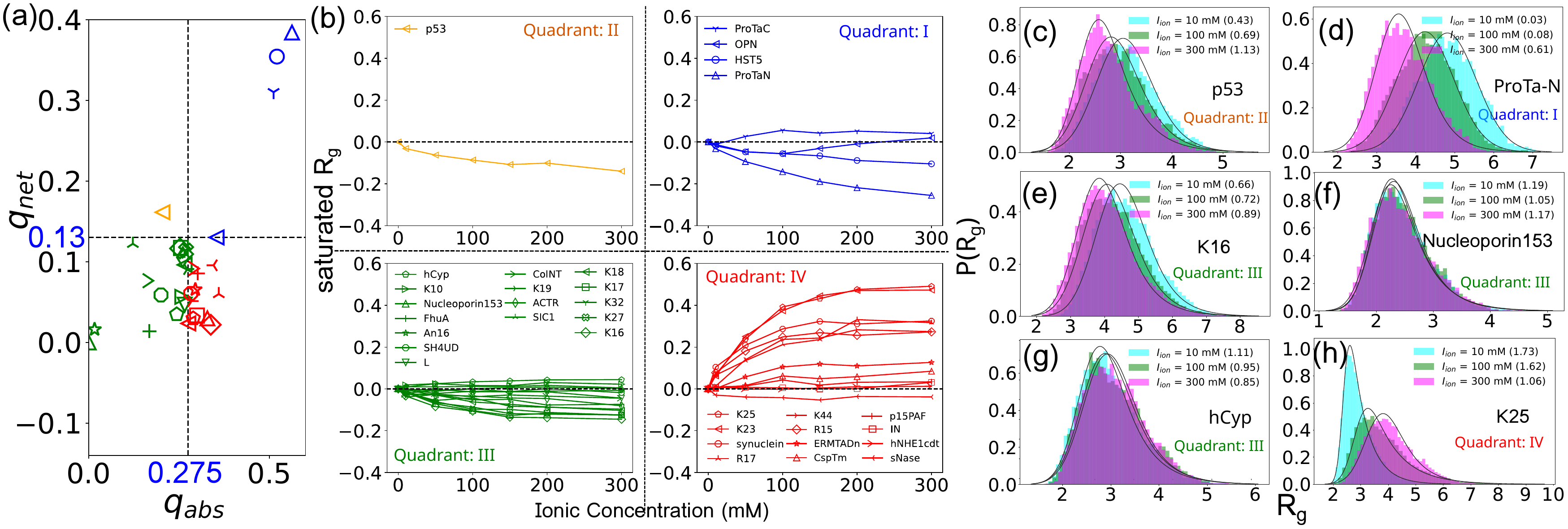}
\caption{\small (a) Characterization of IDPs using their net charge per residue $q_{net}$ as a function of the absolute charge per residue $q_{abs}$. The two dotted black horizontal and vertical lines $q_{net}=0.13$, and $q_{abs}=0.275$ respectively,  further identify the IDPs in terms of their placement into one of the four quadrants (I - IV). (b) $\sqrt{R_g^2/R_g^2(0)}-1$ as a function of ionic concentration $I$. Separate colors and in each quadrant with different symbols for each IDP clearly identifies different dependencies of the IDPs on the salt concentration. $P\left(\bar{R}_g\right)$s 
for p53 (c), ProTa-N (d), K16 (e), Nucleoporin153 (f), hCyp (g), and
K25 (h) are shown for three different ionic concentrations 10 mM
(cyan), 100 mM (green) and 300 mM (magenta) respectively. The black
lines denote the exponentially modified Gaussian fit of the histograms
and the corresponding
$\mathcal{S}$-indices are denoted on the legends.}
\label{i_quads}
\end{figure*}
origin of life. Previous experimental studies
~\cite{Ueda2010,Schuler2010,Hoffmann} and simulation studies using CG
models~\cite{Reddy,Wohl} have revealed 
conformational changes and salt-induced phase transition and looked at
the liquid-liquid phase transitions in IDPs. IDPs are described either
as PEs or PAs with varying amounts of net charge~\cite{Pappu2010}. Thus, it is
conceivable that screening will affect the conformational aspects in a
significant way. Intuitively one can understand the behavior by using the idea of screening. The IDPs that are PE, an increase in salt
concentration will screen the net charge reducing the electrostatic
repulsion and hence by and large, all the PEs with a net positive or 
negative charge will have reduced gyration radii as a function of
increased screening. The case of PAs is a bit more subtle depending on an IDP's not only the net charge per residue $q_{net}$, but the fraction of the residues that are charged $q_{abs}$
as defined below.
\begin{equation}
  q_{net} = \frac{||Q_{+}|-|Q_{-}||}{N} \quad {\rm and} \quad q_{abs} = \frac{|Q_{+}|+|Q_{-}|}{N}
\label{q_definition}
\end{equation} 
Here, $Q_{+}$, $Q_{-}$, and $N$ represent the total positive and
negative charges, and the number of amino acids in the IDP. In
literature, $q_{net}$ and $q_{abs}$ are previously denoted as NCPR and FCR respectively~\cite{Pappu2013}. For the PA the loss/gain in electrostatic energy and entropy ultimately controls the show.
We have made an extensive study of the dependence of gyration radii of the 33 IDPs (listed in Table-~\ref{Table}) on salt concentration under physiological conditions ranging from 0-300 mM  shown in Fig.~\ref{i_quads}. The IDPs can be placed on any one of the four quadrants (I, II, III, and IV) of $(q_{abs}, q_{net})$  to study their dependency on salt concentration (Fig.~\ref{i_quads}(a)).
Based on the values of $(q_{abs}, q_{net})$, two decision boundary lines $q_{abs}=0.275$ and $q_{net}=0.13$ place the IDPs into four subclasses.  Fig.~\ref{i_quads}(a) displays scatter plots of the 33 IDPs classified into four quadrants, represented by blue, green, orange, and red symbols corresponding to the I, II, III, and IV quadrants, respectively. For each quadrant,
a plot of the saturation values at each concentration is used to plot $\sqrt{R_g^2/R_g^2(0)}-1$ as a function of ionic concentration, where $R_g(0)$ corresponds to the radius of gyration under ion-free conditions.
Quadrant-I and IV are easy to understand. 
A strong PA, such as Prota-N lies in the quadrant-I as expected. Other
PAs (HST5, OPN, ProTa-C) with large $q_{net}$ and $q_{abs}$ belong
here. In this case, as the salt concentration increases, charge
screening occurs, leading to a decrease in their gyration radii. This condition holds when only one type of charged residue is abundant in number.
On the other hand, in quadrant IV, $q_{abs}$ is high, but $q_{net}$ is low, corresponding to a situation where there is a higher number of charged residues, yet they are almost equal in numbers.
As both types of charges are present, at low salt concentration, the attraction between opposite charges reduces their radius of gyration due to electrostatic interactions. However, with increasing salt concentration, the charge screening effect comes into play, and the strength of electrostatic attraction among the oppositely charged residues decreases. Consequently, we observe a swelling of the IDPs, leading to a higher radius of gyration. Twelve IDPs belong to this category.
In quadrant II and III, the $q_{abs}$ value is low, indicating a low content of charged residues. In quadrant II, we find that p53 is the only IDP out of the 33 that falls into this category but has a high value of $q_{net}$. This pathological case is characterized by having only 17 negative charge residues (GLU and ASP) and two positively charged residues (ARG and LYS). Due to the charge screening effect mostly on the negatively charged residues, it can be inferred that the radius of gyration will decrease and that is indeed true as observed from the plot.
In quadrant III, the $q_{net}$ is low corresponding to IDPs that have
less net charge per residue. We find 15 IDPs belong to this
category. In this case gyration radii of hCyp, FhuA, and K10 increase
while gyration radii of K18, K17, K32, k27 and K16 decrease as a function of salt concentration. On the other hand, the remaining 5 IDPs, namely Protein-L, SH4UD, An16, Nucleoporin153, and CoINT are robust to the variation of ionic concentration as they are mostly low charge containing IDPs. 
The segregation of IDPs into four subsections unravels insights about their responses to salt concentration and provides a framework to classify other unknown proteins based on how they will behave under a wide range of salt solutions.
\par
We further studied the accompanying variation in the shape of the distribution of the gyration radii by monitoring the $\mathcal{S}$-index as progressively more screening is introduced for the reason discussed in section D. Some examples are shown Fig.~\ref{i_quads}(c)-(h).  The skewness of the distributions for K25 for 10 mM, 100 mM, and 300 mM ionic concentration changes from 1.73, 1.62, and 1.06 respectively, and they span a larger conformational space. On the other hand, for ProTa-N  while the gyration radii decrease at higher salt concentrations implying they become more compact without altering the distribution shape. We also observe that gyration radii for a few IDPs 
(Protein-L, SH4UD, An16, Nucleoporin153, and CoINT) in Fig.~\ref{i_quads}(a) remain unaffected within the low salt limit of our study. With the change in the salt concentration, our simulation studies show that the degree of alteration in the shape of IDPs is different. The shape deformation is drastic in K25 compared to ProTa-N.
\subsection{Summary and Outlook}
In conclusion, we used two different CG models (HPS1 \& HPS2) to study both universal and fine structures of 33 IDPs and compared our results with available experimental data as well as simulation results for the same IDPs using other CG models. Our systematic  studies of IDPs with fairly disparate levels of absolute and net charge 
($q_{abs}$, $q_{net}$), and net hydropathy add many interesting
characteristics to the prior studies using similar models. For the
HPS1 model use of a larger
set of IDPs and new experimental data~\cite{Best_expt} converge on the interaction parameter $\epsilon \approx
0.1$~kcal/mol, as opposed to 0.2 ~kcal/mol if we use the original set
of Dignon {\em et al.}~\cite{Mittal2018} (please refer to the
supplementary section). For the HPS2 model, our optimized parameter
0.2 ~kcal/mol using 33 IDPs
concurs with that of Tesei {\em et al.}~\cite{Larsen2021,Tesai_Calvados} and seem to be more robust
compared to the HPS1
model.  Analysis of the results using the expanded
dataset establishes a more robust and reliable framework for studying
IDPs in bulk.
The experimental results converge well with the simulation results for the optimum $\epsilon$ for both HPS1 and HPS2 models. \par
A natural question that has been addressed in the
community that if sequence specificity makes every IDP distinct from
each other, or they share some universal characteristics of
homopolymers described by Flory's theory. We have been able to address
both issues. A comparison of the scaled end-to-end distance and
transverse fluctuations in reference to our recently established
universal results~\cite{Universal1}, we observe that IDPs studied here
interpolates from being Gaussian chains  to self-avoiding chains to a
different degree depending upon their amino acid composition. 
\par
We then study in detail how the absolute and net charge per residue ($q_{abs}$, $q_{net}$) manifest themselves in finer characteristics of the IDPs. We come up with several new metrics that reveal the uniqueness of each IDP, yet leaves room for making further classification of IDPs in different categories. The first one is the Wilson index $\mathcal{W}$ that on a normalized unit length scale demonstrates the uniqueness of each IDP
and hence can be used as their fingerprints ({\em e.g.} We believe that using the
$\mathcal{W}$ index of the newly discovered IDPs in future will allow us
to predict their behaviors with those of similar $\mathcal{W}$ index. \par
Likewise, we demonstrate how the charge patches control the local
stiffness and hence the overall conformations of the IDPs that we
further characterize by introducing a skewness index $\mathcal{S}$.
We first demonstrate that the distribution of the scaled gyration radii for highly charged IDPs with $\mathcal{S} \rightarrow 0 $ are near-perfect Gaussians
and that moderately charged IDPs with a larger value of
the  $\mathcal{S} \rightarrow 1.0 - 1.5$ are characterized by 
exponential tails. This observation
immediately leads us to the observation of a simple linear dependence of  $\mathcal{S}$ on
$q_{net}$  as approximated in Eqn.~\ref{S-qnet}. We related  
$\mathcal{S}$ parameter to the net charge charge asymmetry parameter $f^*$ introduced
by Das and Pappu~\cite{Pappu2013} that we observe a stronger
exponential dependence. We believe
Fig.~\ref{skewness-fit} along with Eqn.~\ref{S-qnet} can serve as a good reference to understand properties
of the individual IDPs.

\par
An important classification of the IDPs emerges from the study of salt dependence of the conformations of the IDPs. We find that IDPs exhibit very different characteristics and can be broadly placed in four different region in the ($q_{abs}, q_{net}$) space.\par
We conclude with some comments which may promote further studies to
perfect the CG models. We and many others used isotropic radius of
gyration as the sole physical quantity for comparison as this is the
mostly available from the experiments.  The CG model can be refined by
introducing other quantities.  For example, Wohl {\em et al.}~\cite{Wohl} studied  salting-out effect on the liquid–liquid phase separation (LLPS) of IDPs by introducing a salt-dependent term into the hydropathy used in the HPS1 model.
Maity {\em et al.}~\cite{Reddy} introduced Molecular Transfer Model to study the Salt-Induced Transitions. In the low concentrations of salts ($\le$ 1M) IDP conformations are affected by the degree of screening of electrostatic interactions of the charged residues and are independent of the specific salt identity which is likely the regime that we have studied. However, at high concentrations, salts affect IDP conformations through salt-specific Hofmeister effects~\cite{Capp2013,Pegram2010}. Thus, our studies will be useful in refining the existing hydropathy models for a wider range of parameter space.
\par
We note that some of the IDPs remain unaffected with the variation of
salt concentration and thus can be compared with the behavior of other
simpler amino acids identified and studied in the context of the
origin of life~\cite{Origin_life}. Recently Tesei {\em et al.}
developed an efficient model to generate conformational ensembles
of IDRs, and reported conformational properties 28,058 IDRs from sequence only~\cite{Tesei2023}.
We believe that some of the new ideas introduced here can be tested
for a much larger set and will open up several exciting avenues for
future research to obtain a deeper understanding of the unique properties and behavior of IDPs.\\
\section{Supplementary Materials}
A comparison of the gyration radii data from our simulation based on
HPS1 hydropathy scale with the experimental data for a smaller subset of IDPs ~\cite{Mittal2018} is included in the supplementary materials.
\section{Acknowledgments}
All computations were carried out using STOKES High-Performance
Computing Cluster at UCF. We thank Prof. Kresten Lindorff-Larsen and
Dr. Giulio Tesei from University of Copenhagen for discussions,
comparison of simulation results,  and bringing reference
~\cite{Best_expt} to our attention. We sincerely thank both the referees for their critical comments on
the manuscript. Simulation results were obtained using HOOMD-blue~\cite{HOOMD}.
\vfill


\begin{thebibliography}{100}

\bibitem{Uversky2000} Uversky, V. N., Gillespie, J. R. \& Fink, A. L. Why are ``natively unfolded" proteins unstructured under physiologic conditions? Proteins: Structure, Function, and Genetics 41, 415-427 (2000).

\bibitem{Dunker} Oldfield, C. J. \& Dunker, A. K. Intrinsically Disordered Proteins and Intrinsically Disordered Protein Regions. Annual Review of Biochemistry 83, 553-584 (2014).

\bibitem{Best-COSB-2017} Best, R. B. Computational and theoretical advances in studies of intrinsically disordered proteins. Current Opinion in Structural Biology 42, 147-154 (2017).

\bibitem{IDP_Review1} Ghosh, K., Huihui, J., Phillips, M. \& Haider, A. Rules of Physical Mathematics Govern Intrinsically Disordered Proteins. Annual Review of Biophysics 51, 355-376 (2022). 

\bibitem{IDP_Review2} Ehm, T. {\em et al.} Intrinsically disordered proteins at the nano-scale. Nano Futures 5, 022501 (2021).

\bibitem{IDP_Review3} Evans, R., Ramisetty, S., Kulkarni, P. \& Weninger, K. Illuminating Intrinsically Disordered Proteins with Integrative Structural Biology. Biomolecules 13, 124 (2023).

\bibitem{Pappu2014}van der Lee, R. {\em et al.} Classification of Intrinsically Disordered Regions and Proteins. Chemical Reviews 114, 6589-6631 (2014).

\bibitem{DisProt} Sickmeier, M. {\em et al.} DisProt: the Database of Disordered Proteins. Nucleic Acids Research 35, D786-D793 (2007).

\bibitem{Giansanti} Deiana, A., Forcelloni, S., Porrello, A. \& Giansanti, A. Intrinsically disordered proteins and structured proteins with intrinsically disordered regions have different functional roles in the cell. PLOS ONE 14, e0217889 (2019).

\bibitem{IDP_cell_signalling} Wright, P. E. \& Dyson, H. J. Intrinsically disordered proteins in cellular signalling and regulation. Nature Reviews Molecular Cell Biology 16, 18-29 (2014).

\bibitem{Ferrie} Ferrie, J. J., Karr, J. P., Tjian, R. \& Darzacq, X. ``Structure''-function relationships in eukaryotic transcription factors: The role of intrinsically disordered regions in gene regulation. Molecular Cell 82, 3970-3984 (2022).

\bibitem{Best_Nature2018} Borgia, A. {\em et al.} Extreme disorder in an ultrahigh-affinity protein complex. Nature 555, 61-66 (2018).

\bibitem{Schuler_NatCommn} Sottini, A. {\em et al.} Polyelectrolyte interactions enable rapid association and dissociation in high-affinity disordered protein complexes. Nature Communications 11, 5736 (2020).
  
\bibitem{Fung2018} Fung, H. Y. J., Birol, M. \& Rhoades, E. IDPs in macromolecular complexes: the roles of multivalent interactions in diverse assemblies. Current Opinion in Structural Biology 49, 36-43 (2018).

\bibitem{IDP_Review4} Uversky, V. N., Oldfield, C. J. and Dunker, A. K. Intrinsically Disordered Proteins in Human Diseases: Introducing the $\rm{D_2}$ Concept. Annual Review of Biophysics 37, 215-246 (2008).

\bibitem{Uversky2022} Coskuner-Weber, O., Mirzanli, O. \& Uversky, V. N. Intrinsically disordered proteins and proteins with intrinsically disordered regions in neurodegenerative diseases. Biophysical Reviews 14, 679-707 (2022).

\bibitem{Pappu_NatPhys} Brangwynne, C. P., Tompa, P. \& Pappu, R. V. Polymer physics of intracellular phase transitions. Nature Physics 11, 899-904 (2015).

\bibitem{Mittal_Nature} Murthy, A. C. {\em et al.} Molecular interactions underlying liquid-liquid phase separation of the FUS low-complexity domain. Nature Structural Molecular Biology 26, 637-648 (2019).
  
\bibitem{Chan_PNAS} Das, S., Lin, Y.-H., Vernon, R. M., Forman-Kay, J. D. \& Chan, H. S. Comparative roles of charge, $\pi$, and hydrophobic interactions in sequence-dependent phase separation of intrinsically disordered proteins. Proceedings of the National Academy of Sciences 117, 28795-28805 (2020).

\bibitem{Pappu2010} Mao, A. H., Crick, S. L., Vitalis, A., Chicoine, C. L. \& Pappu, R. V. Net charge per residue modulates conformational ensembles of intrinsically disordered proteins. Proceedings of the National Academy of Sciences 107, 8183-8188 (2010). 

\bibitem{Pappu2013} Das, R. K. \& Pappu, R. V. Conformations of intrinsically disordered proteins are influenced by linear sequence distributions of oppositely charged residues. Proceedings of the National Academy of Sciences 110, 13392-13397 (2013). 

\bibitem{Schuler_PNAS2012} Hofmann, H. {\em et al.} Polymer scaling laws of unfolded and intrinsically disordered proteins quantified with single-molecule spectroscopy. Proceedings of the National Academy of Sciences 109, 16155-16160 (2012).


\bibitem{Svergun} Bernadó, P. \& Svergun, D. I. Structural analysis of intrinsically disordered proteins by small-angle X-ray scattering. Mol. BioSyst. 8, 151-167 (2012).

\bibitem{Schuler_Review} Schuler, B., Soranno, A., Hofmann, H. \& Nettels, D. Single-Molecule FRET Spectroscopy and the Polymer Physics of Unfolded and Intrinsically Disordered Proteins. Annual Review of Biophysics 45, 207-231 (2016).

\bibitem{Schuler_JCP2018} Schuler, B. Perspective: Chain dynamics of unfolded and intrinsically disordered proteins from nanosecond fluorescence correlation spectroscopy combined with single-molecule FRET. The Journal of Chemical Physics 149, 010901 (2018).

\bibitem{Best_FRET} Merchant, K. A., Best, R. B., Louis, J. M., Gopich, I. V. \& Eaton, W. A. Characterizing the unfolded states of proteins using single-molecule FRET spectroscopy and molecular simulations. Proceedings of the National Academy of Sciences 104, 1528-1533 (2007).

\bibitem{Tompa2013} Kosol, S., Contreras-Martos, S., Cedeño, C. \& Tompa, P. Structural Characterization of Intrinsically Disordered Proteins by NMR Spectroscopy. Molecules 18, 10802-10828 (2013).


\bibitem{Gomes_SIC1} Gomes, G.-N. W. {\em et al.} Conformational Ensembles of an Intrinsically Disordered Protein Consistent with NMR, SAXS, and Single-Molecule FRET. Journal of the American Chemical Society 142, 15697-15710 (2020).


\bibitem{Sokal} Sokal, A. D. In Monte Carlo and Molecular Dynamics Simulations in Polymer Science; Binder, K., Ed.; Oxford University Press: New York, 1995; Chapter 2.

\bibitem{Ausbaugh} Ashbaugh, H. S. \& Hatch, H. W. Natively Unfolded Protein Stability as a Coil-to-Globule Transition in Charge/Hydropathy Space. Journal of the American Chemical Society 130, 9536-9542 (2008).

\bibitem{Mittal2018} Dignon, G. L., Zheng, W., Kim, Y. C., Best, R. B. \& Mittal, J. Sequence determinants of protein phase behavior from a coarse-grained model. PLOS Computational Biology 14, e1005941 (2018).

\bibitem{Larsen2021} Tesei, G., Schulze, T. K., Crehuet, R. \& Lindorff-Larsen, K. Accurate model of liquid-liquid phase behavior of intrinsically disordered proteins from optimization of single-chain properties. Proceedings of the National Academy of Sciences 118, e2111696118 (2021).

\bibitem{Pappu_Package} Lalmansingh, J. M., Keeley, A. T., Ruff, K. M., Pappu, R. V. \& Holehouse, A. S. SOURSOP: A Python Package for the Analysis of Simulations of Intrinsically Disordered Proteins. Journal of Chemical Theory and Computation 19, 5609-5620 (2023).


\bibitem{Thirumalai_2019} Baul, U., Chakraborty, D., Mugnai, M. L., Straub, J. E. \& Thirumalai, D. Sequence Effects on Size, Shape, and Structural Heterogeneity in Intrinsically Disordered Proteins. The Journal of Physical Chemistry B 123, 3462-3474 (2019).

\bibitem{Thirumalai_2023} Mugnai, M. L. {\em et al.} Sizes, conformational fluctuations, and SAXS profiles for Intrinsically Disordered Proteins. (2023) doi:10.1101/2023.04.24.538147.
 
\bibitem{Best_expt} Dannenhoﬀer-Lafage, T., \& Best, R, B. A Data-Driven Hydrophobicity Scale for Predicting Liquid-Liquid Phase Separation of Proteins. J. Phys. Chem. B 125, 4046-4056 (2021).
  
\bibitem{COINT} Johnson, C. L. {\em et al.} The Two-State Prehensile Tail of the Antibacterial Toxin Colicin N. Biophysical Journal 113, 1673-1684 (2017).
  
\bibitem{FhuA} Riback, J. A. {\em et al.} Innovative scattering analysis shows that hydrophobic disordered proteins are expanded in water. Science 358, 238–241 (2017).
  
  
\bibitem{OPN} Platzer, G. {\em et al.} The Metastasis-Associated Extracellular Matrix Protein Osteopontin Forms Transient Structure in Ligand Interaction Sites. Biochemistry 50, 6113-6124 (2011).
  
\bibitem{histatin5} Jephthah, S., Staby, L., Kragelund, B. B. \& Skep\"{o}, M. Temperature Dependence of Intrinsically Disordered Proteins in Simulations: What are We Missing? Journal of Chemical Theory and Computation 15, 2672-2683 (2019).

\bibitem{Kyte} Kyte, J. \& Doolittle, R. F. A simple method for displaying the hydropathic character of a protein. Journal of Molecular Biology 157, 105-132 (1982).

\bibitem{Habchi} Habchi, J., Tompa, P., Longhi, S. \& Uversky, V. N. Introducing Protein Intrinsic Disorder. Chemical Reviews 114, 6561-6588 (2014).

\bibitem{Mittal2022} Devarajan, D. S. {\em et al.} Effect of Charge Distribution on the Dynamics of Polyampholytic Disordered Proteins. Macromolecules 55, 8987-8997 (2022).

\bibitem{Alberti-LLPS2019} Alberti, S., Gladfelter, A. \& Mittag, T. Considerations and Challenges in Studying Liquid-Liquid Phase Separation and Biomolecular Condensates. Cell 176, 419-434 (2019).

\bibitem{Dorfmann-LLPS2019} Alberti, S. \& Dormann, D. Liquid-Liquid Phase Separation in Disease. Annual Review of Genetics 53, 171-194 (2019).

\bibitem{McCarty-LLPS2019} McCarty, J., Delaney, K. T., Danielsen, S. P. O., Fredrickson, G. H. \& Shea, J.-E. Complete Phase Diagram for Liquid-Liquid Phase Separation of Intrinsically Disordered Proteins. The Journal of Physical Chemistry Letters 10, 1644-1652 (2019).

\bibitem{Muthukumar_MM2022} Das, S. \& Muthukumar, M. Microstructural Organization in $\alpha$-Synuclein Solutions. Macromolecules 55, 4228-4236 (2022). 

\bibitem{Aksimentiev_JPCL2020} Chou, H.-Y. \& Aksimentiev, A. Single-Protein Collapse Determines Phase Equilibria of a Biological Condensate. The Journal of Physical Chemistry Letters 11, 4923-4929 (2020).

\bibitem{Israel} J. N. Israelachvili, {\em Intermolecular and Surface forces}, 3rd edition, Elsevier (2011). 

\bibitem{Akerlof} Akerlof, G. C. \& Oshry, H. I. The Dielectric Constant of Water at High Temperatures and in Equilibrium with its Vapor. Journal of the American Chemical Society 72, 2844-2847 (1950).

\bibitem{Engelman} Engelman, D. M., Steitz, T. A. \& Goldman, A. Identifying nonpolar transbilayer helices in amino acid sequences of membrane proteins. Annual Review of Biophysics and Biophysical Chemistry 15, 321-353 (1986).

\bibitem{Hopp} Hopp, T. P. \& Woods, K. R. A computer program for predicting protein antigenic determinants. Molecular Immunology 20, 483-489 (1983).

\bibitem{Eisenberg} Eisenberg, D., Schwarz, E., Komaromy, M. \& Wall, R. Analysis of membrane and surface protein sequences with the hydrophobic moment plot. Journal of Molecular Biology 179, 125-142 (1984).

\bibitem{Cornette} Cornette, J. L. {\em et al.} Hydrophobicity scales and computational techniques for detecting amphipathic structures in proteins. Journal of Molecular Biology 195, 659-685 (1987).

\bibitem{Everaers_PRL} Yamakov, V., Milchev, A., J\"{o}rg Limbach, H., D\"{u}nweg, B. \& Everaers, R. Conformations of Random Polyampholytes. Physical Review Letters 85, 4305-4308 (2000).

\bibitem{Pincus_MM_1980} Schaefer, D. W., Joanny, J. F. \& Pincus, P. Dynamics of Semiflexible Polymers in Solution. Macromolecules 13, 1280-1289 (1980). 

\bibitem{Nakanishi_1987}
Nakanishi, H. Flory approach for polymers in the stiff limit. Journal de Physique 48, 979-984 (1987).

\bibitem{Rubinstein} M. Rubinstein and R.H. Colby, {\it Polymer Physics} (Oxford Univ. Press, 2003). 

\bibitem{Universal1} Bair, J., Seth, S. \& Bhattacharya, A. Universality in conformations and transverse fluctuations of a semi-flexible polymer in a crowded environment. The Journal of Chemical Physics 158, 204902 (2023).

\bibitem{Universal2} Huang, A., Bhattacharya, A. \& Binder, K. Conformations, transverse fluctuations, and crossover dynamics of a semi-flexible chain in two dimensions. The Journal of Chemical Physics 140, 214902 (2014).

\bibitem{SK-Ma} S.~K. Ma, {\em Modern Theory of Critical Phenomena}, (Taylor and Francis, USA, 1976).
  
\bibitem{Landau} L. D. Landau and E. M. Lifshitz, Statistical Physics, Part 1, 3rd ed. (Pergamon Press, 1980).

\bibitem{exponnorm} scipy.stats.exponnorm — SciPy v1.12.0.dev Manual. (n.d.). URL: http://scipy.github.io/devdocs \\
/reference/generated/scipy.stats.exponnorm.html

\bibitem{Ueda2010} Matsunaga, H. \& Ueda, H. Stress-induced non-vesicular release of prothymosin-$\alpha$ initiated by an interaction with S100A13, and its blockade by caspase-3 cleavage. Cell Death \& Differentiation 17, 1760-1772 (2010).

\bibitem{Schuler2010} M\"{u}ller-Sp\"{a}th, S. {\em et al.} Charge interactions can dominate the dimensions of intrinsically disordered proteins. Proceedings of the National Academy of Sciences 107, 14609-14614 (2010).

\bibitem{Hoffmann} Vancraenenbroeck, R., Harel, Y. S., Zheng, W. \& Hofmann, H. Polymer effects modulate binding affinities in disordered proteins. Proceedings of the National Academy of Sciences 116, 19506-19512 (2019).

\bibitem{Reddy} Maity, H., Baidya, L. \& Reddy, G. Salt-Induced Transitions in the Conformational Ensembles of Intrinsically Disordered Proteins. The Journal of Physical Chemistry B 126, 5959-5971 (2022).

\bibitem{Wohl} Wohl, S., Jakubowski, M. \& Zheng, W. Salt-Dependent Conformational Changes of Intrinsically Disordered Proteins. The Journal of Physical Chemistry Letters 12, 6684-6691 (2021).

\bibitem{Tesai_Calvados} Tesei, G and Lindorff-Larsen K., Improved 
  predictions of phase behaviour of intrinsically disordered proteins 
  by tuning the interaction range, bioRxiv preprint doi: 
  https://doi.org/10.1101/2022.07.09.499434 (July 13, 2022). 
  
  
\bibitem{Capp2013} Record, M. T., Guinn, E., Pegram, L. \& Capp, M. Introductory Lecture: Interpreting and predicting Hofmeister salt ion and solute effects on biopolymer and model processes using the solute partitioning model. Faraday Discuss. 160, 9-44 (2013).

\bibitem{Pegram2010} Pegram, L. M. {\em et al.} Why Hofmeister effects of many salts favor protein folding but not DNA helix formation. Proceedings of the National Academy of Sciences 107, 7716-7721 (2010).

\bibitem{Origin_life} Pohorille, A., Wilson, M. A. \& Shannon, G. Flexible Proteins at the Origin of Life. Life 7, 23 (2017). 
  
\bibitem{Tesei2023} Tesei, G. {\em et al.} Conformational ensembles of
  the human intrinsically disordered proteome: Bridging chain
  compaction with function and sequence conservation. bioRxiv 2023.05.08.539815 (2023) doi:
   https://doi.org/10.1101/2023.05.08.539815.
   

\bibitem{HOOMD}Anderson, J. A., Glaser, J., \& Glotzer, S. C. HOOMD-blue: A Python package for high-performance molecular dynamics and hard particle Monte Carlo simulations. Computational Materials Science 173: 109363, Feb 2020. 10.1016/j.commatsci.2019.109363

\end{thebibliography}
\end{document}